\documentclass[10pt,conference]{IEEEtran}
\IEEEoverridecommandlockouts
\usepackage{cite}
\usepackage{amsmath,amssymb,amsfonts}
\usepackage{algorithmic}
\usepackage{paralist} 
\usepackage{graphicx}
\usepackage[belowskip=-2pt,aboveskip=2.5pt,font=small]{caption}
\usepackage{subcaption}
\usepackage{url}
\usepackage{array}
\usepackage{textcomp}
\usepackage{multirow}
\usepackage{enumitem}
\usepackage{flushend}
\usepackage{xcolor}
\newcolumntype{L}[1]{>{\raggedright\let\newline\\\arraybackslash\hspace{0pt}}m{#1}}
\newcolumntype{C}[1]{>{\centering\let\newline\\\arraybackslash\hspace{0pt}}m{#1}}
\newcolumntype{R}[1]{>{\raggedleft\let\newline\\\arraybackslash\hspace{0pt}}m{#1}}

\def\BibTeX{{\rm B\kern-.05em{\sc i\kern-.025em b}\kern-.08em
    T\kern-.1667em\lower.7ex\hbox{E}\kern-.125emX}}
\begin{document}

\title{Automating App Review Response Generation
}

\author{\IEEEauthorblockN{Cuiyun Gao\IEEEauthorrefmark{2}~~~Jichuan Zeng\IEEEauthorrefmark{2}~~~Xin Xia\IEEEauthorrefmark{3}~~~David Lo\IEEEauthorrefmark{4}~~~Michael R. Lyu\IEEEauthorrefmark{2}~~~Irwin King\IEEEauthorrefmark{2}}
\IEEEauthorblockA{\IEEEauthorrefmark{2}Department of Computer Science and Engineering, The Chinese University of Hong Kong, Hong Kong, China\\\IEEEauthorrefmark{3}Faculty of Information Technology, Monash University, Australia\\\IEEEauthorrefmark{4}School of Information Systems, Singapore Management University, Singapore}
\{cygao,jczeng,lyu,king\}@cse.cuhk.edu.hk~~~xin.xia@monash.edu~~~davidlo@smu.edu.sg}


\maketitle

\begin{abstract}
Previous studies showed that replying to a user review usually has a positive effect on the rating that is given by the user to the app. For example, Hassan et al. found that responding to a review increases the chances of a user updating their given rating by up to six times compared to not responding. To alleviate the labor burden in replying to the bulk of user reviews, developers usually adopt a template-based strategy where the templates can express appreciation for using the app or mention the company email address for users to follow up. However, reading a large number of user reviews every day is not an easy task for developers. Thus, there is a need for more automation to help developers respond to user reviews.


Addressing the aforementioned need, in this work we propose a novel approach RRGen that automatically generates review responses by learning knowledge relations between reviews and their responses. RRGen explicitly incorporates review attributes, such as user rating and review length, and learns the relations between reviews and corresponding responses in a supervised way from the available training data. Experiments on 58 apps and 309,246 review-response pairs highlight that RRGen outperforms the baselines by at least 67.4\% in terms of BLEU-4 (an accuracy measure that is widely used to evaluate dialogue response generation systems). Qualitative analysis also confirms the effectiveness of RRGen in generating relevant and accurate responses.


\end{abstract}


\section{Introduction}\label{sec:introduction}
Mobile apps are software applications designed to run on smartphones, tablets and other mobile devices. They already serve as an integral part of people's daily life, and continuously gain traction over the last few years. The apps are typically available from app stores, such as Apple's App Store and Google Play. These app stores allow users to express their opinions to apps by writing reviews and giving ratings. User experience determines if users will keep using an app or uninstall it, possibly posting favorable or unfavorable feedbacks. For example, a survey in 2015~\cite{apptentive} reported that 65\% users chose to leave a rating or review after a negative experience, and only 15\% users would consider downloading an app with a 2-star rating. To compete with the bulk of the apps offering similar functionalities, ensuring good user experience is crucial for app developers.

App reviews act as one direct communication channel between developers and users, delivering users' instant experience after their interactions with apps. Analysis on app reviews can assist developers in discovering in a timely manner important app issues, such as bugs to fix or requested features, for app maintenance and development~\cite{cygao2018idea,di2016would}. Currently, both Apple's App Store and Google Play provide a review response system for developers to manually respond to a review, after which the corresponding user who posted the review will be notified and have the option to update their reviews~\cite{iosresponse,googleresponse}. In the response, developers can talk about the roadmap about users' proposed feature requests, explain the usage of app functionalities, or just thank users for their shared opinions.

Empirical studies~\cite{DBLP:conf/chi/OhKLLS13, DBLP:journals/software/McIlroySAH17, DBLP:journals/ese/HassanTBH18,DBLP:journals/corr/abs-1808-03796} that analyze the interactions between users and developers demonstrate that responding to user feedback in a timely and accurate manner can (1) enhance app development and (2) improve user experience. Specifically, Nayebi et al.~\cite{DBLP:journals/corr/abs-1808-03796} automatically summarized user requests which was proven to shorten the cycle between issue escalation and developers' fix. McIlroy et al.~\cite{DBLP:journals/software/McIlroySAH17} observed that users change their rating 38.7\% of the time following a developer response. Hassan et al.~\cite{DBLP:journals/ese/HassanTBH18} found that developers of 34.1\% of the apps they analyzed respond to at least one review, and also confirmed the positive effect of the responses on rating change. For example, they discovered that the number of users who increases their ratings after receiving a response are six times more than those who receive no response. App developers can also solve 34\% of the reported issues without deploying an update. In spite of the benefits of the review-response mechanism, due to the large and ever-increasing number of reviews received daily, many reviews still did not receive timely response~\cite{DBLP:journals/ese/HassanTBH18,iosresponse}. This highlights the necessity and importance of automatic response generation, which is the focus of our work.



Dialogue generation has been extensively studied in the natural language processing field~\cite{DBLP:conf/emnlp/WangJBN17, DBLP:conf/emnlp/LiS18, DBLP:conf/naacl/SordoniGABJMNGD15}, for facilitating social conversations, e.g., the Microsoft XiaoIce chatbot~\cite{DBLP:journals/corr/abs-1812-08989}. Such work is generally grounded in the basic RNN Encoder-Decoder model (or Neural Machine Translation model, abbreviated as NMT)~\cite{DBLP:conf/emnlp/ChoMGBBSB14, DBLP:journals/corr/SutskeverVL14}, where the context and corresponding response are regarded as source and target sequences respectively. The RNN Encoder-Decoder model is an end-to-end learning approach for automated translation. It has been applied to a number of software engineering tasks, such as producing a sequence of APIs given a natural language query~\cite{DBLP:conf/sigsoft/GuZZK16}, parsing natural language into machine interpretable sequences (e.g., database queries)~\cite{DBLP:conf/acl/DongL16}, generating commit messages according to code changes~\cite{DBLP:conf/iwpc/JiangM17,DBLP:conf/kbse/JiangAM17}, and inferring variable types based on contextual code snippets~\cite{DBLP:conf/sigsoft/HellendoornBBA18}. However, the applicability of the NMT model for app review response generation has not been studied. To fill in this gap, we explore the usability of the NMT model in the app review-response dialogue scenario here, by regarding user reviews and the corresponding replies as the source and target sequences respectively.

Directly applying the NMT model to app dialogue generation may not be appropriate, since the app review-response dialogues and social conversations are different in many ways. First, the purpose of app dialogues is to further understand users' complaints or solve user requests, while social conversations are mainly for entertainment purpose. This implies that app reviews require more accurate and clearer response~\cite{iosresponse}. Second, users' sentiment expressed in reviews should be precisely identified. Although reviews contain the information of star ratings, the ratings and actual emotions may not be totally consistent~\cite{islam2014numeric,DBLP:conf/ACMse/SharmaL13}. For example, one user may write positive feedback like ``\textit{Great}'', but only give one-star rating. Third, app reviews are generally short in length and usually with only one round of dialogue. According to Hassen et al.~\cite{DBLP:journals/ese/HassanTBH18}, 97.5\% of the app dialogues end after one iteration. Such limited context increases the difficulty of generating a concise response.

In this paper, we propose an improved NMT model, named RRGen, for accurate \textbf{R}eview \textbf{R}esponse \textbf{Gen}eration. We extend basic NMT by incorporating review-specific characteristics (e.g., star ratings and review lengths) to capture user's sentiment and complaint topics. To evaluate the effectiveness of our model, we collected 309,246 review-response pairs from 58 popular apps published on Google Play. For a comprehensive comparison, besides the basic NMT model, we also choose the state-of-the-art approach in commit message generation based on code changes~\cite{DBLP:conf/kbse/LiuXHLXW18}, named NNGen, as one baseline model. Because NNGen adopts basic information retrieval technique which is commonly used in traditional dialogue generation tasks~\cite{DBLP:journals/corr/JiLL14,DBLP:conf/ijcai/SongLNZZY18,DBLP:books/daglib/0021593}, and claims better performance than the basic NMT model. Our experimental results show that RRGen significantly outperforms the baseline models by 67.4\%$\sim$450\% in terms of BLEU-4 score~\cite{DBLP:conf/acl/PapineniRWZ02} (an accuracy measure that is widely used to evaluate dialogue response generation systems). Human evaluation done through a user study also indicates that RRGen can generate a more relevant and accurate response than NNGen. Besides reporting the promising results, we investigate the reason behind the superior performance of our model and the key constraints on automatic response generation.

The main contributions of our work are as follows:
\begin{itemize}
\item To our knowledge, we are the first to consider the problem of automatic review response generation, and propose a deep neural network technique for solving the problem. We propose a novel neural machine translation model, RRGen\footnote{available at: \url{https://github.com/ReMine-Lab/RRGen}}, to learn both topics and sentiments of reviews for a accurate response generation.
\item The accuracy of RRGen is empirically evaluated using a corpus of more than 300 thousand real-life review-response pairs. A user study was also conducted to verify RRGen's effectiveness in generating reasonable reviews. 
\end{itemize}

\textbf{Paper structure.} Section~\ref{sec:back} illustrates the background of review-response system, and neural encoder-decoder model. Section~\ref{sec:model} presents our proposed model for user review response generation. Section~\ref{sec:data} and Section~\ref{sec:exper} describe our experimental setup and the quantitative evaluation results. Section~\ref{sec:human} details the results of a human evaluation of our proposed model. Section~\ref{sec:dis} discusses the advantages, limitation, and threats of our work. Related work and final remarks are discussed in Section~\ref{sec:related} and Section~\ref{sec:con}, respectively.


\section{Background}\label{sec:back}
Our work adopts and augments advanced techniques from deep learning and neural machine translations~\cite{DBLP:journals/corr/BahdanauCB14,DBLP:conf/icml/CollobertB04,DBLP:conf/emnlp/LuongPM15}. In this section, we introduce the user-developer dialogue and discuss the background of these techniques.

\subsection{User-Developer Dialogue}
\begin{figure}[h]
	\centering
	\includegraphics[width=0.48 \textwidth]{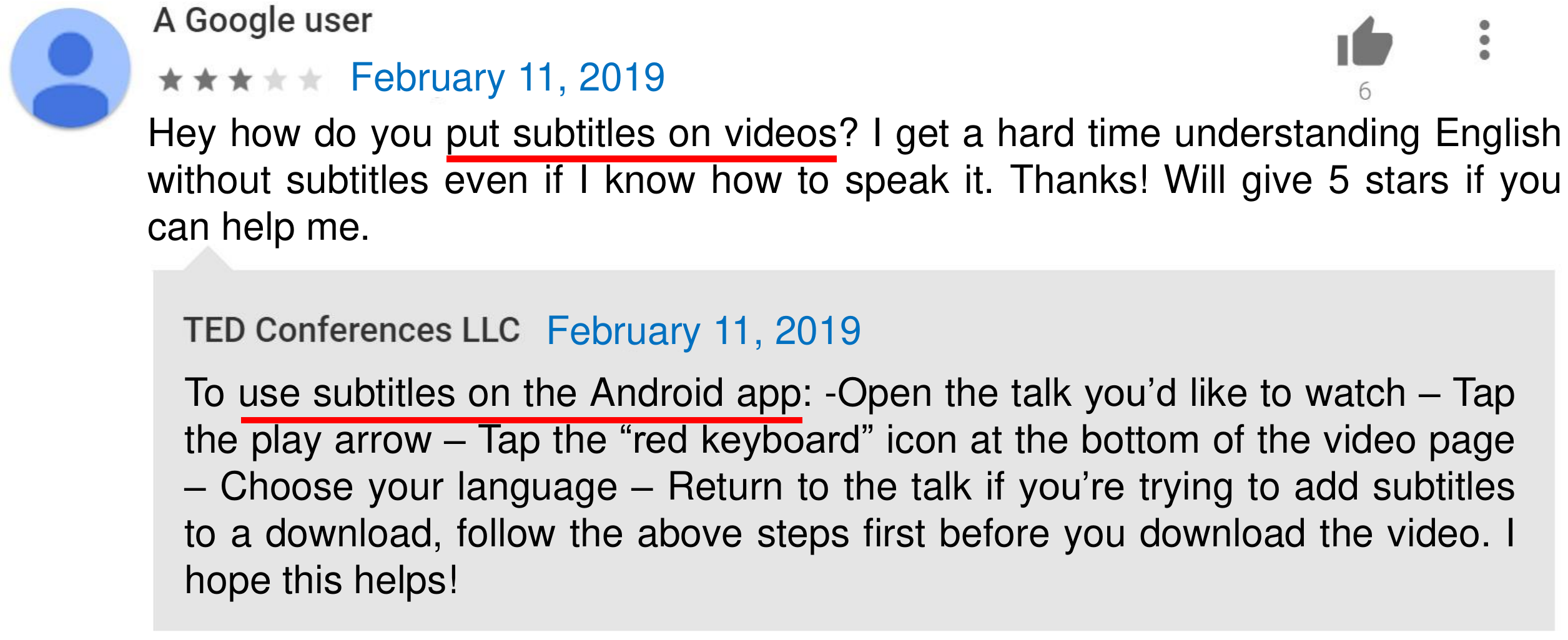}
	\caption{Example of TED developer's response to one user review. The red underlines highlight some topical words of the dialogue.}
	\label{fig:reply_example}
\end{figure}

Figure~\ref{fig:reply_example} depicts an example of the user-developer dialogue of the TED app in Google Play. A user initiates the dialogue by posting a review, including a star rating, for an app. User reviews convey valuable information to developers, such as major bugs, feature requests, and simple complaints or praise about the experience~\cite{DBLP:conf/re/MaalejN15}. As encouraged by the App Store~\cite{iosresponse}, responding to feedback in a timely and consistent manner can improve user experience and an app's ranking. For example, the review in Fig.~\ref{fig:reply_example} was complaining about the unclear functionality usage related to adding ``\textit{video subtitles}''. The TED developer then responded with detailed steps for putting subtitles, and later, the user changed the star rating to five. 

Generally, developers could not reply to all app reviews due to their limited time and efforts, and also a large number of reviews. As studied by Hassan et al.~\cite{DBLP:journals/ese/HassanTBH18}, developers respond to 2.8\% of the collected user reviews, and they tend to reply reviews with low ratings and long contents. The App Store also suggests developers to consider prioritizing reviews with the lowest star ratings or those mentioning technical issues for responding~\cite{iosresponse}. However, ranking reviews for developers' reply is out of the scope of this work, and the related studies can be found in~\cite{DBLP:journals/software/McIlroySAH17,DBLP:journals/ese/HassanTBH18}. We focus on alleviating the manual labor in responding to feedback and aim at automating the process. Moreover, since 97.5\% of the app dialogues end after one round~\cite{DBLP:journals/ese/HassanTBH18}, in this study, we concentrate on one iteration of user review reply.

\subsection{RNN Encoder-Decoder Model}
\begin{figure}[h]
	\centering
	\includegraphics[width=0.35 \textwidth]{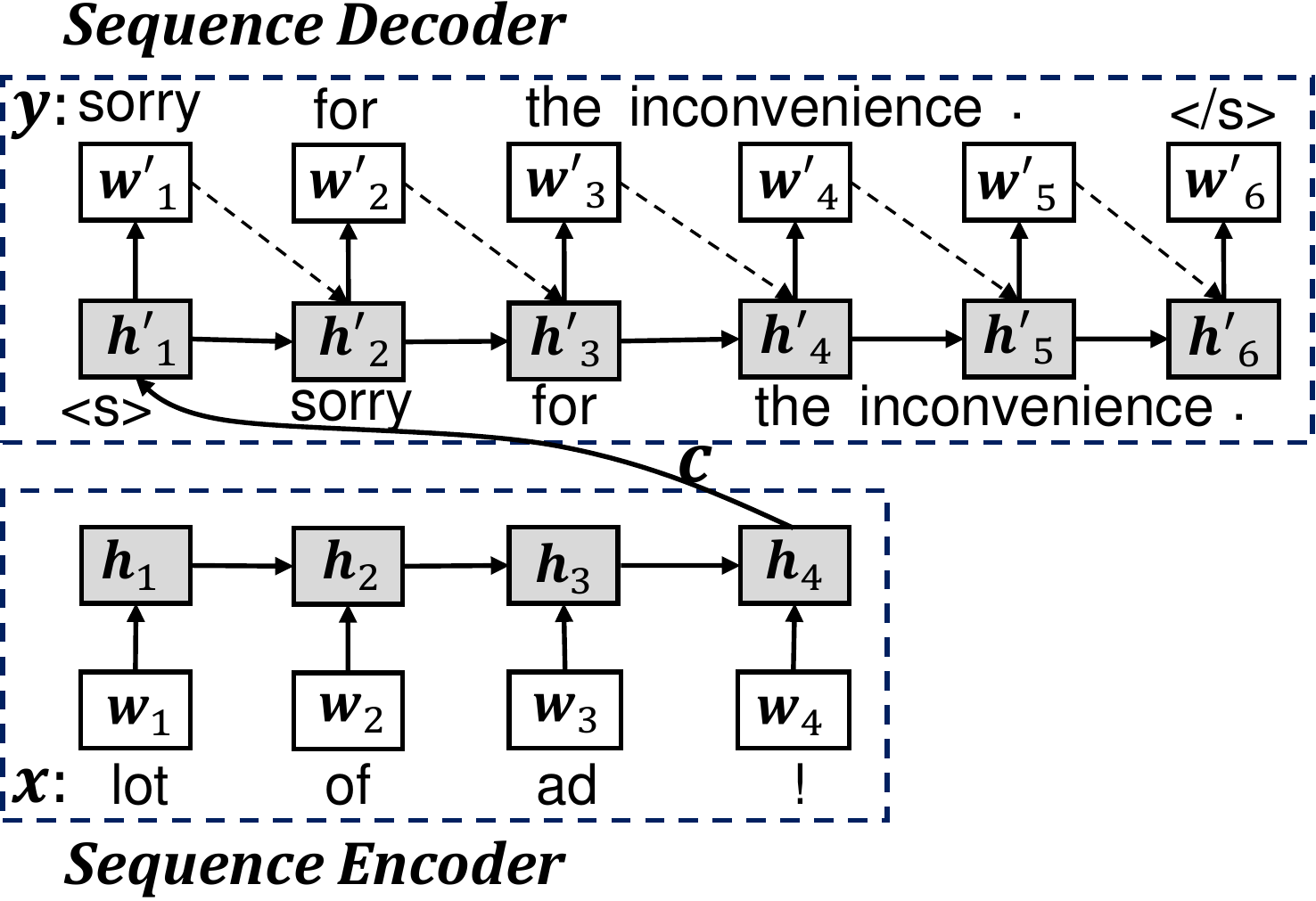}
	\caption{An overall architecture of RNN encoder-decoder model.}
	\label{fig:rnn}
\end{figure}

The RNN Encoder-Decoder~\cite{DBLP:conf/emnlp/ChoMGBBSB14} model is an effective and standard approach for neural machine translation and sequence-to-sequence (seq2seq)~\cite{DBLP:conf/nips/SutskeverVL14} prediction. In general, the RNN encoder-decoder models aim at generating a target sequence $\boldsymbol{y}=(y_1,y_2,...,y_{T_y})$ given a source sequence $\boldsymbol{x}=(x_1,x_2,...,x_{T_x})$, where $T_x$ and $T_y$ are sequence lengths of the source and target respectively. Fig.~\ref{fig:rnn} illustrates an overall architecture of the RNN encoder-decoder model.

To do so, an encoder first converts the source sequence $\boldsymbol{x}$ into a set of hidden vectors $\{\boldsymbol{h}_1,\boldsymbol{h}_2,...,\boldsymbol{h}_{T_x}\}$, whose size varies regarding the source sequence length. The context representation $\boldsymbol{c}$ is generated using a Recurrent Neural Network (RNN)~\cite{DBLP:conf/interspeech/MikolovKBCK10}. The encoder RNN reads the source sentences from the first token until the last one, where $\boldsymbol{h}_t=f(\boldsymbol{h}_{t-1}, \boldsymbol{w}_t)$, and $\boldsymbol{c}=\boldsymbol{h}_{T_x}$. Here, $\boldsymbol{w}_t$ is the word embedding of the source token $x_t$, where word embeddings~\cite{DBLP:conf/nips/MikolovSCCD13} are distributed representations of words in a continuous vector space, and trained with a text corpus. The $f$ is a non-linear function that maps a the word embedding $\boldsymbol{w}_t$ into a hidden state $\boldsymbol{h}_t$ by considering the previous hidden state $\boldsymbol{h}_{t-1}$.

Then, the decoder, which is also implemented as an RNN, generates one word $y_t$ at each time stamp $t$ based on the hidden state $\boldsymbol{h}_t^\prime$ as well as the previous predicted word $y_{t-1}$:

\begin{equation}
Pr(y_t|y_{t-1},...,y_1,\boldsymbol{c})=g(\boldsymbol{h}_t^\prime,y_{t-1},\boldsymbol{c}),
\end{equation}

\noindent where $g$ is a non-linear mapping function, and the context vector $\boldsymbol{c}$ returned by the encoder is set as an initial hidden state, i.e., $\boldsymbol{h}_1^\prime=\boldsymbol{c}$. The decoder stops when generating the end-of-sequence word \textit{\textless\textbackslash s\textgreater}.

The two RNN encoder-decoder models are jointly trained to maximize the conditional log-likelihood:
\begin{equation}\label{equ:cross}
\mathcal{L}(\theta) = \max_\theta\frac{1}{N}\sum_{i=1}^{N}\log p_\theta(\boldsymbol{y}_i|\boldsymbol{x}_i),
\end{equation}
where $\theta$ is the set of the model parameters (e.g., weights in the neural network) and each $(\boldsymbol{x}_i,\boldsymbol{y}_i)$ is a (source sequence, target sequence) pair from the training set. The $p_\theta (\boldsymbol{y}_i|\boldsymbol{x}_i)$ denotes the likelihood of generating the $i$-th target sequence $\boldsymbol{y}_i$ given the source sequence $\boldsymbol{x}_i$ according to the model parameters $\theta$. Through optimizing the loss function using optimization algorithms such as gradient descent, the optimum $\theta$ values can be estimated.

\subsection{Attention Mechanism}\label{subsec:att}
\begin{figure}[h]
	\centering
	\includegraphics[width=0.35 \textwidth]{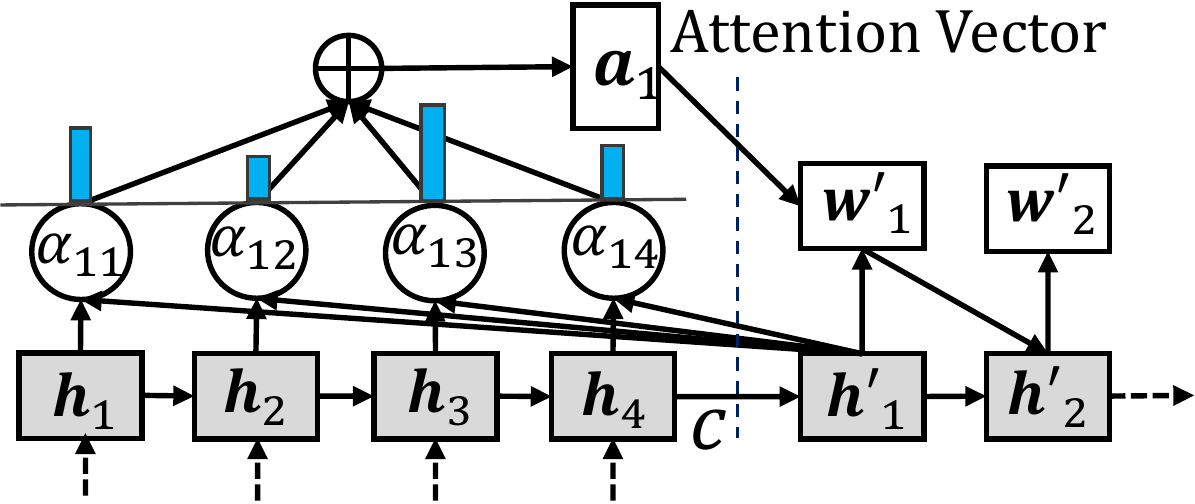}
	\caption{Graphical illustration of the attentional RNN encoder-decoder model. The dotted line without arrow marks the division between the encoder (left) and decoder (right), and the dotted lines with arrows indicate that we simplify the RNN encoder-decoder~\cite{DBLP:conf/emnlp/ChoMGBBSB14} steps for clearness.}
	\label{fig:attention}
\end{figure}

A potential issue with the RNN encoder-decoder model is that a neural network needs to compress all the necessary information of a source sequence into a fixed-length vector. To alleviate this issue, Bahdanau et al.~\cite{DBLP:journals/corr/BahdanauCB14} proposed the attention mechanism to focus on relevant parts of the source sequence during decoding. We use the attention mechanism in our work because previous studies~\cite{DBLP:conf/emnlp/RushCW15,DBLP:journals/corr/LinFSYXZB17,DBLP:conf/emnlp/ZengLSGLK18} prove that attention-based models can better capture the key information (e.g., topical or emotional tokens) in the source sequence. Fig.~\ref{fig:attention} shows a graphical illustration of the attentional RNN encoder-decoder model.

During decoding, besides the hidden state $\boldsymbol{h}_t^\prime$ and previous predicted word $y_{t-1}$, an attention vector $\boldsymbol{a}_t$ is also involved for generating one word $y_t$ at each time stamp $t$:
\begin{equation}
Pr(y_t|y_{t-1},...,y_1,\boldsymbol{c})=g(\boldsymbol{h}_t^\prime,y_{t-1},\boldsymbol{c},\boldsymbol{a}_t).
\end{equation}
The attention vector $\boldsymbol{a}_t$ depends on the relevance between the hidden state $\boldsymbol{h}_t^\prime$ and the encoded source sequence $(\boldsymbol{h}_1,...,\boldsymbol{h}_{T_x})$:
\begin{equation}\label{equ:attention}
\boldsymbol{a}_t = \sum_{j=1}^{T_x}{\alpha_{tj}\boldsymbol{h}_j},
\end{equation}
\noindent where $T_x$ is the length of the source sequence, and the attention weight $\alpha_{tj}$ measures how helpful the $j$-th hidden state of the source sequence $\boldsymbol{h}_j$ is in predicting next word $y_t$ with respect to the previous hidden state $\boldsymbol{h}_{t-1}^\prime$. In this way, the decoder decides parts of the source sentence to pay attention to.

\begin{figure*}[ht]

        \begin{subfigure}[h]{0.66\textwidth}
        \centering
    	\includegraphics[width=0.97 \textwidth]{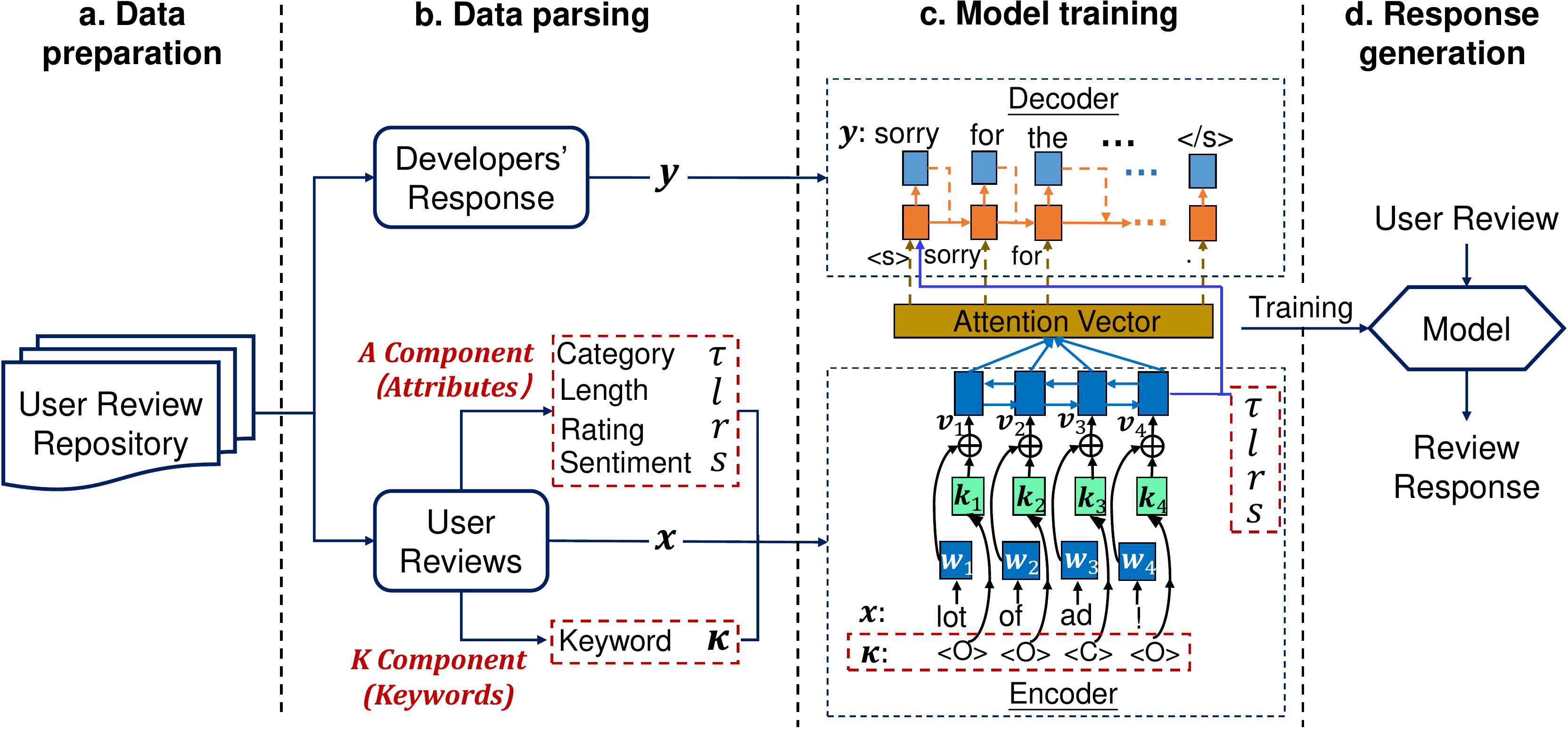}
    	\caption{\small{Overall architecture of RRGen}}
        \end{subfigure}
        \begin{subfigure}[h]{0.34\textwidth}
        \centering
        \includegraphics[width=0.85 \textwidth]{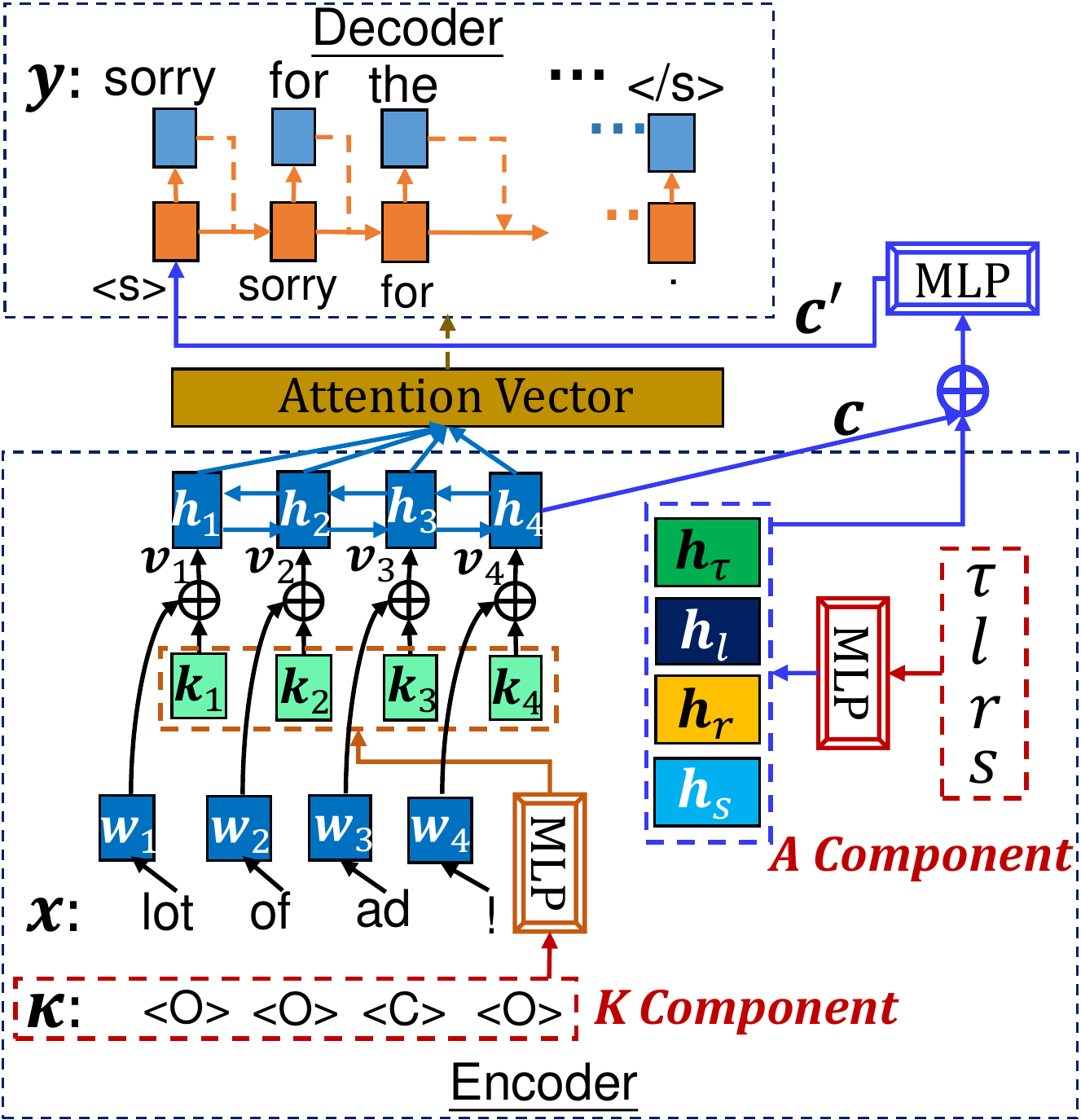}
    	\caption{\small Detailed structure of RRGen}
        \end{subfigure}
    \caption{Structure of the review response generative model.}
	\label{fig:overall}
\end{figure*}

\section{RRGen: App Review Response Generation}\label{sec:model}
In this section, we present the design of RRGen that extends the basic attentional RNN Encoder-Decoder model for app review response generation. We regard user reviews as the source sequence and developers' response as the target sequence. Fig.~\ref{fig:rnn} shows an example of the RNN Encoder-Decoder model for generating a sequence of tokens as a developer's response from a sequence of tokens that constitute a user review ``\textit{Lot of ad!}''. For accurately capturing the topics and sentiment embedded in the input review sequence, we explicitly incorporate both high-level attributes (e.g., app category, review length, user rating, and sentiment) and keywords into the original RNN Encoder-Decoder model. We adopt the keywords provided by Di Sorbo et al.~\cite{di2016would} which were manually curated to identify 12 topics (e.g., GUI, contents, pricing, etc.) commonly covered in user reviews. We refer to the high-level attributes and keywords extracted from a review as its A and K components, respectively.

Figure~\ref{fig:overall} (a) shows the overall architecture of our RRGen model. RRGen mainly consists of four stages: Data preparation, data parsing, model training, and response generation. We first collect app reviews and their responses from Google Play, and conduct preprocessing. The preprocessed data are parsed into a parallel corpus of user reviews and their corresponding responses, during which the two components of reviews are also extracted and processed. Based on the parallel corpus of app reviews and responses, we build and train a generative neural model with the two pieces of extracted information (high-level attributes and keywords) holistically considered. The major challenge during the training process lies in the effective consideration of both components of reviews for effective response generation. In the following, we will introduce the details of the RRGen model and the approach we propose to resolve the challenge.



\subsection{Component Incorporation}
Here, we elaborate on how we incorporate the two components, including high-level attributes (or A Component) and keywords (or K Component), into RRGen. The detailed structure of RRGen is displayed in Fig.~\ref{fig:overall} (b).

\subsubsection{A Component} The A component contains four attributes of one user review: App category, review length, user rating, and sentiment. We choose app category considering that apps of different categories generally contain different functionalities, and major topics delivered by their reviews would be different. Review length is involved because it is an important index of whether the review is informative or not, i.e., longer reviews usually convey richer information~\cite{chen2014ar,DBLP:journals/ese/HassanTBH18}. We take user rating into account since it can directly impact the response style of developers, e.g., expressing an apology for negative feedback or thanks for the positive feedback. As user ratings may not be consistent with the sentiment described by the reviews~\cite{guzman2014users}, we also regard the predicted actual user sentiment as one attribute. 

Review attributes such as app category, review length, and user rating are easy to acquire. For predicting user sentiment, we exploit SentiStrength~\cite{DBLP:journals/jasis/ThelwallBPCK10}, a lexical sentiment extraction tool specialized in handling short and low-quality texts. We first divide review text into sentences, and then assigns a positive integer value (in the range [+1, +5]) and a negative integer value (within the range [-5, -1]) based on StentiStrength to each sentence because users may express both positive and negative sentiments in the same sentence. A higher absolute sentiment score indicates that the corresponding sentiment is stronger. Following Guzman and Maalej's work~\cite{guzman2014users}, when the sentence's negative score multiplied by 1.5 is less than the positive score, we assign the sentence a negative sentiment score; Otherwise, the sentence is assigned a positive sentiment score~\cite{guzman2014users}. The sentiment of an entire review is computed based on the rounded average sentiment scores of all sentences in the review.

We denote the app category, review length, user rating, and sentiment score of the source sequence $\boldsymbol{x}$ as $\tau$, $l$, $r$, and $s$, respectively. To incorporate these attributes into RRGen, we first represent the attribute values into continuous vectors via multilayer perceptions (MLPs), i.e., the conventional fully connected layer~\cite{DBLP:conf/ijcai/MontanaD89}. We call the vector representations of the attributes as attribute embeddings. The embedding of app category $\tau$ is defined as:

\begin{equation}
    \boldsymbol{h}_{\tau} = \tanh(\boldsymbol{W}^\Gamma \operatorname{Emb}(\tau)), \forall \tau=1,2,...,N_\Gamma,
\end{equation}

\noindent where $\boldsymbol{W}^\Gamma$ is the matrix of trainable parameters in the MLP, and $\boldsymbol{h}_{\tau}, g=1,...,N_\Gamma$ are the embedding vectors of all individual categories. $\operatorname{Emb}(\tau) \in \mathbb{R}^{N_\Gamma}$ is the vector representation of $\tau$, and $\operatorname{Emb}(\cdot)$ indicates one general embedding layer to obtain the latent features of $\tau$. Similarly, we obtain the embedding vectors for user rating $r$ and sentiment score $s$:




\begin{equation}
    \boldsymbol{h}_{r} = \tanh(\boldsymbol{W}^R \operatorname{Emb}(r)), \forall r=1,2,...,N_R,
\end{equation}
\begin{equation}
    \boldsymbol{h}_{s} = \tanh(\boldsymbol{W}^S \operatorname{Emb}(s)), \forall s=1,2,...,N_S,
\end{equation}

\noindent where $\boldsymbol{h}_r$ and $\boldsymbol{h}_s$ are embeddings for the attribute values $r$ and $s$, respectively. For review length $l$, we convert the continuous variable into its categorical form $l^\prime$ using the \textit{pandas} package\footnote{\url{https://pandas.pydata.org/pandas-docs/stable/}} before feeding into MLP.

\begin{equation}
    \boldsymbol{h}_{l} = \tanh(\boldsymbol{W}^L \operatorname{Emb}(l^\prime), \forall l^\prime=1,2,...,N_L.
\end{equation}

We integrate the embedded attribute values at review level by concatenating together with the last hidden state $\boldsymbol{c}$ of the encoder, i.e., 
\begin{equation}
    \boldsymbol{c}^\prime = \tanh(\boldsymbol{W}^H[\boldsymbol{c};\boldsymbol{h}_{\tau};\boldsymbol{h}_l;\boldsymbol{h}_r;\boldsymbol{h}_s]),
\end{equation}

\noindent where $[\boldsymbol{a};\boldsymbol{b}]$ is the concatenation of these two vectors. $\boldsymbol{W}^H$ is the matrix of trainable parameters in the MLP, and $H$ is the number of hidden units. The output vector $\boldsymbol{c}^\prime$ indicates the final hidden state (or context vector) of the encoder. For simplicity, we assume that the dimensions of all attribute embeddings, i.e., $\boldsymbol{h}_\tau$, $\boldsymbol{h}_l$, $\boldsymbol{h}_r$, and $\boldsymbol{h}_s$, are the same.


\subsubsection{K Component}
K component specifically refers to keywords in the input review sequence, since the keywords generally relate to the review topic or sentiment, and are potentially helpful to learn which word to attend to during response generation.

\begin{table}[h]
	\center
	\caption{One example of topic-keywords pair in the keyword dictionary provided by Di Sorbo et al.~\cite{di2016would}.}
	\label{tab:keyword}
	\scalebox{1.0}{\begin{tabular}{l| L{7.2cm}}
		\hline
		 \textbf{Topic} & \textbf{Keywords} \\
		 \hline
        GUI & screen, trajectory, button, white, background, interface, usability, tap, switch, icon, orientation, picture, show, list, category, cover, scroll, touch, clink, snap, underside, backside, witness, rotation, ui, gui,... \\
		\hline
	\end{tabular}
}
\end{table}


We adopt the keyword dictionary provided by Di Sorbo et al.~\cite{di2016would}. Di Sorbo et al. summarize 12 topics\footnote{The 12 topics are app, GUI, contents, pricing, feature, improvement, updates/versions, resources, security, download, model, and company.} commonly covered by user reviews based on manual analysis, and build a keyword dictionary based on WordNet~\cite{DBLP:journals/cacm/Miller95} to extract related words for each topic. One topic-keywords pair can be seen in Table~\ref{tab:keyword}. Di Sorbo et al. utilized the dictionary to predict topics of user reviews and achieved $>$90\% classification accuracy; this indicates the semantic representativeness of these keywords for each topic. This motivates us to use the keywords too in our work. 

To explicitly integrate the keyword information into RRGen, we establish a keyword sequence $\boldsymbol{\kappa}=(\kappa_1,\kappa_2,...,\kappa_{T_x})$ for each input review sequence $\boldsymbol{x}$. Specifically, for the token $x_t$ in $\boldsymbol{x}$, we check the keyword dictionary to determine its subordinate topic, i.e., $\kappa_t$. For example, as shown in Fig.~\ref{fig:overall} (b), the keyword sequence corresponding to the source sequence ``\textit{lot of ad !}'' is ``\textless O\textgreater \textless O\textgreater \textless C\textgreater \textless O\textgreater'', where we denote the keyword symbol for the token ``\textit{ad}'' as ``\textless C\textgreater'' since ``\textit{ad}'' is one keyword for topic \textit{contents}. The keyword symbols of non-topical words (e.g., ``\textit{of}'') are labeled as ``\textless O\textgreater''. We finally integrate the embedded keyword sequence and the source sequence at token level via MLP:

\begin{equation}
    \begin{array}{c}
        \boldsymbol{k}_\iota = \tanh(\boldsymbol{W}^K \operatorname{Emb}(\kappa_\iota)) , \forall \iota=1,2,...,N_K \\
        \boldsymbol{v}_t = \tanh(\boldsymbol{W}^V [\boldsymbol{k}_t;\boldsymbol{w}_t])
    \end{array}
\end{equation}
\noindent where $\boldsymbol{k}_\iota, \iota=1,...,N_K$ are the embedding vectors of all individual keyword symbols, $\boldsymbol{W}^K$ and $\boldsymbol{W}^V$ are the matrices of trainable parameters in the MLPs, and $\boldsymbol{v}_t$ is the keyword-enhanced embedding for the $t$-th token $x_t$ in the source sequence. The dimension of $\boldsymbol{k}_\iota$ is similar to the attribute embeddings in the A component, e.g, $\boldsymbol{h}_\tau$.



\subsection{Model Training and Testing}
\subsubsection{Training}
We adopt the attention mechanism, described in Section~\ref{subsec:att}, for review response generation. The RNN has various implementations, we use bidirectional Gated Recurrent Units (GRUs)~\cite{DBLP:conf/emnlp/ChoMGBBSB14} which is a popular RNN encoder-decoder model and performs well in many tasks~\cite{DBLP:journals/corr/ChungGCB14,DBLP:conf/icassp/WuK16}. All GRUs have 200 hidden units in each direction. Each attribute in the two components is encoded into an embedding with dimension at 90, i.e., the embedding size of $\boldsymbol{h}_\tau, \boldsymbol{h}_l, \boldsymbol{h}_r, \boldsymbol{h}_s$, and $\boldsymbol{k}_\iota$. Word embeddings are initiated with pre-trained 100-dimensional GloVe vectors~\cite{glove}. We set the maximum sequence length at 200 and save the model every 200 batches. We discuss the details of parameter tuning in Section~\ref{subsec:param}. The training goal is cross-entropy minimization based on Equ.~(\ref{equ:new_cross}):

\begin{equation}\label{equ:new_cross}
\mathcal{L}(\theta) = \max_\theta\frac{1}{N}\sum_{i=1}^{N}\log p_\theta(\boldsymbol{y}_i|\boldsymbol{x}_i,\tau, l, r, s, \boldsymbol{\kappa}_i),
\end{equation}

\noindent where $\tau, l, r, s, \boldsymbol{\kappa}_i$ correspond to the app category, review length, user rating, sentiment score, and keyword sequence of the $i$-th source sequence $\boldsymbol{x}_i$, respectively. The whole model is trained using the minibatch Adam~\cite{DBLP:journals/corr/KingmaB14}, a stochastic optimization approach and automatically adjusting the learning rate. We set the batch size (i.e., number of review instances per batch) as 32. For training the neural networks, we limit the source and target vocabulary to the top 10,000 words that are most frequently used in user reviews and developers' responses.

For implementation, we use PyTorch~\cite{pytorch}, an open-source deep learning framework. We train our model in a server with one Nvidia TITAN V GPU with 12GB memory. The training lasts $\sim$80 hours with two epochs.

\subsubsection{Testing}
We evaluate on the test set when the trained model after one batch shows an improvement on the validation set regarding BLEU score~\cite{DBLP:conf/acl/PapineniRWZ02}. We take the highest test score and corresponding generated response as the evaluation result. We use the same GPU as we used in training. The testing process took around 25 minutes.

\section{Experimental Setup}\label{sec:data}

\subsection{Data Preparation}
\subsubsection{Data Collection}
We select the subject apps for collecting the user-developer dialogues from Google Play based on app popularity. We focus on popular apps since they contain more reviews than unpopular apps~\cite{DBLP:conf/msr/HarmanJZ12}, which should facilitate enough data for studying user-developer dialogues. We select the top 100 free apps in 2016 according to App Annie~\cite{appannie}, an app analytics platform, as these apps were top apps two years prior to the start of our study. The decision was made to ensure the studied apps had enough reviews to collect and also avoid the influence of an app's price on developers' review response behavior~\cite{DBLP:journals/ese/HassanTBH18}. We further remove the apps that are no longer available in Google Play on April 2018 and those with fewer than 100 user reviews, which leaves us with 72 apps that match our selection criteria.

For each selected app, we created a Google Play crawler to collect user-developer dialogues from Google Play, specifically including review title, review text, review post time, user name, rating, developer response time, and the text in the developer response. We run our crawler from April 2016 to April 2018. During that period, we collected 15,963,612 reviews for the 72 apps. We find that 58/72 apps and 318,973 collected reviews have received a response from the app developer. Table~\ref{tab:dataset} describes the statistics of the 58 subject apps which belong to 15 app categories.

\subsubsection{Data Preprocessing}
Since app reviews are generally submitted via mobile terminals and written using limited keyboards, they contain massive noisy words, such as repetitive words and misspelled words~\cite{cygao2018idea}. We first convert all the words in the reviews and their response into lowercase, and adopt the method in~\cite{DBLP:conf/issre/ManGLJ16} for lemmatization. We then replace all digits with ``\textless digit\textgreater''. We also detect email address and URL with regular expressions, and substitute them into ``\textless email\textgreater'' and ``\textless url\textgreater'' respectively. Besides, we build an app list containing all the app names, and a user list with all the user names. For the app names and user names mentioned in the dialogue corpus, we replace them with ``\textless app\textgreater'' and ``\textless user\textgreater'' respectively. We finally adopt the rule-based methods based on~\cite{DBLP:conf/kbse/VuNPN15,DBLP:conf/issre/ManGLJ16} to rectify repetitive words and misspelled words. After removing empty review texts or review texts with only one single alphabet, we obtained 309,246 review-response pairs. We randomly split the dataset by 8:1:1, as the training, validation, and test sets, i.e.,  there are 279,792, 14,727, and 14,727 pairs in the training, validation, and test sets, respectively.

\begin{table}[h]
	\center
	\caption{Mean and five-number summary of collected data for every studied app.}
	\label{tab:dataset}
	\scalebox{0.85}{\begin{tabular}{l r r r r r r}
		\hline
		  & Avg. & Min. & 1st Qu. & Med. & 3rd Qu. & Max.\\
		 \hline
        \#reviews per app & 203,025 & 5,582 & 83,317 & 179,457 & 287,286 & 665,203\\
        \#reviews with &  \multirow{2}{*}{5,406} & \multirow{2}{*}{2} & \multirow{2}{*}{181} & \multirow{2}{*}{1,149} & \multirow{2}{*}{4,290}& \multirow{2}{*}{55,165}\\
        responses per app & & &  &  &  & \\
		\hline
	\end{tabular}
}
\end{table}
\vspace{-0.4cm}



\subsection{Similarity Measure - BLEU}
BLEU~\cite{DBLP:conf/acl/PapineniRWZ02} is a standard automatic metric for evaluating dialogue response generation systems. It analyzes the co-occurrences of $n$-grams in the ground truth $\boldsymbol{y}$ and the generated responses $\hat{\boldsymbol{y}}$, where $n$ can be 1, 2, 3, or 4. BLEU-$N$, where $N$ is the maximum length of $n$-grams considered, measures the proportion of co-occurrences of $n$ consecutive tokens between the ground truth $\boldsymbol{y}$ and generated response $\hat{\boldsymbol{y}}$. The most commonly used version of BLEU uses $N=4$~\cite{DBLP:conf/kbse/LiuXHLXW18,DBLP:conf/kbse/JiangAM17}, i.e., BLEU-4. Also, BLEU-4 is usually calculated at the corpus-level, which is demonstrated to be more correlated with human judgments than other evaluation metrics~\cite{DBLP:conf/emnlp/LiuLSNCP16}. Thus, we use corpus-level BLEU-4 as our evaluation metric.





\subsection{Baseline Approaches}
We compare the performance of our model with a random selection approach, the basic attentional RNN encoder-decoder (NMT) model~\cite{DBLP:journals/corr/BahdanauCB14} (as introduced in Section~\ref{subsec:att}), and a state-of-the-art approach for code commit message generation~\cite{DBLP:conf/kbse/LiuXHLXW18}, namely NNGen. In the following, we elaborate on the first and last baselines:

\textit{\textbf{Random Selection:}} This is a strawman baseline. This baseline randomly picks a response in the training set and uses it as a response to a review in the test set.

\textit{\textbf{NNGen:}} We choose NNGen as one comparing approach since it is demonstrated to perform better than the basic NMT model~\cite{DBLP:conf/kbse/JiangAM17} in producing code commit message based on code changes. NNGen leverages the nearest neighbor (NN) algorithm to retrieve the most relevant developer response. Based on the training set and the new user review, NNGen first represents them as vectors in the form of ``bags of words''~\cite{DBLP:books/daglib/0021593}, and then selects the top five training user reviews which present highest cosine similarities to the new review. After that, the BLEU-4 score between the new review and each of the top five training reviews is computed. NNGen finally regards the response of the training review with the highest BLEU-4 score as the result.

\section{Evaluation Using An Automatic Metric}\label{sec:exper}
In this section, we conduct quantitative analysis to evaluate the effectiveness of RRGen. In particular, we intend to answer the following research questions.

\begin{enumerate}[label=\bfseries RQ\arabic*:,leftmargin=.5in]
    \item What is the accuracy of RRGen?
    \item What is the impact of different component attributes on the performance of RRGen?
    \item How accurate is RRGen under different parameter settings?
\end{enumerate}


\subsection{RQ1: What is the accuracy of RRGen?}\label{subsec:accuracy}

The comparison results with baseline approaches are shown in Table~\ref{tab:rq1}. We can see that our RRGen approach outperforms all the three baselines. Specifically, the result that random selection approach achieves the lowest BLEU-4 score (6.55), indicates that learning knowledge from existing review-response pairs can facilitate generating the response for a newly-arrived review. Also, we find that the NMT model performs better than the non-deep-learning-based NNGen model, which shows an increasing rate of 53.48\% in terms of BLEU-4 score. This is opposite to the conclusion achieved by Liu et al.~\cite{DBLP:conf/kbse/LiuXHLXW18}. One possible reason is that the tasks between ours and Liu et al.'s~\cite{DBLP:conf/kbse/LiuXHLXW18} are different, i.e., Liu et al. aim at producing texts based on code, while we focus on generating texts for dialogues and modeling code is different from modeling dialogue texts~\cite{DBLP:conf/acl/YinN17,DBLP:conf/acl/LiuFCRYL18}. The higher BLEU-4 score of the proposed RRGen model than that of the NMT model explains that the response generated by the RRGen model is more similar to developers' response than the response generated by the NMT model. We then use Wilcoxon signed-rank test~\cite{wilcoxon1945individual} for statistical significance test, and Cliff's Delta (or $d$    ) to measure the effect size~\cite{DBLP:journals/technometrics/Ahmed06}. The significance test result ($p-value < 0.01$) and large effect size on BLEU-4 scores ($d=0.74$) of RRGen and NMT confirm the superiority of RRGen over NMT.

\begin{table}[h]
	\center
	\caption{Comparison results with baseline approaches. The $p_n$ indicates the $n$-gram precision when comparing the ground truth and generated responses. Statistical significance results are indicated with \textsuperscript{*}($p-value<0.01$).}
	\label{tab:rq1}
	\scalebox{1.0}{\begin{tabular}{l c c c c c}
		\hline
		 Approach & BLEU-4 & $p_1$ & $p_2$ & $p_3$ & $p_4$\\
		 \hline
        Random & 6.55 & 27.64 & 6.90 & 3.55 & 2.78 \\
        NNGen~\cite{DBLP:conf/kbse/LiuXHLXW18} & 14.08 & 34.47 & 13.85 & 9.77 & 8.59 \\
        NMT~\cite{DBLP:journals/corr/BahdanauCB14} & 21.61 & 40.55 & 20.75 & 16.78 & 15.47 \\\hline
        RRGen & \textbf{36.17\textsuperscript{*}} & \textbf{53.24\textsuperscript{*}} & \textbf{35.83\textsuperscript{*}} & \textbf{31.73\textsuperscript{*}} & \textbf{30.04\textsuperscript{*}} \\
		\hline
	\end{tabular}
}
\end{table}


\subsection{RQ2: What is the impact of different component attributes on the performance of RRGen?}
To evaluate the effectiveness of different component attributes in response generation, we perform contrastive experiments in which only a single component attribute is added to the basic NMT model~\cite{DBLP:journals/corr/BahdanauCB14}. Table~\ref{tab:rq2} shows the results.

Unsurprisingly, the combination of all component attributes gives the highest improvements, and all the attributes are beneficial on their own. User sentiment gives the lowest improvement (+0.58 in terms of BLEU-4 score) comparing to the NMT model, while the app category yields highest improvement (+9.92 in terms of BLEU-4 score). Also, the result that user rating contributed more on the BLEU-4 score than user sentiment indicates that user ratings would be more helpful in review response generation. Moreover, the gain from different component attributes is not fully cumulative since the information encoded in these component attributes overlaps. For instance, both the user sentiment and user rating attributes encode the user emotion expressed by user reviews. Also, the keywords in the K component highlights the words belonging to the same topics, and such information may be already captured by the word embeddings~\cite{DBLP:conf/nips/MikolovSCCD13}. 

\begin{table}[h]
	\center
	\caption{Contrastive experiments with individual component attributes.}
	\label{tab:rq2}
	\scalebox{0.9}{\begin{tabular}{l l c c c c c}
		\hline
		 \multicolumn{2}{c}{Approach} & BLEU-4 & $p_1$ & $p_2$ & $p_3$ & $p_4$\\
		\hline
        \multicolumn{2}{c}{NMT~\cite{DBLP:journals/corr/BahdanauCB14}} & 21.61 & 40.55 & 20.75 & 16.78 & 15.47 \\
        \hline
        \multirow{4}{*}{A Component} & +App Category & 31.53 & 47.49 & 30.64 & 26.84 & 25.30 \\
        & +Review Length & 24.22 & 41.96 & 22.30 & 18.16 & 16.76 \\
        & +Rating & 26.90 & 46.19 & 26.06 & 21.69 & 20.12 \\
        & +Sentiment & 22.19 & 40.42 & 20.95 & 16.99 & 15.69 \\
        \hline
        K Component & +Keyword & 24.34 & 43.41 & 23.66 & 19.27 & 17.74 \\
        \hline
        \multicolumn{2}{c}{RRGen} & \textbf{36.17} & \textbf{53.24} & \textbf{35.83} & \textbf{31.73} & \textbf{30.04} \\
		\hline
	\end{tabular}
}
\end{table}

        

\begin{figure}[h]
    \begin{subfigure}[h]{0.5\textwidth}
    \centering
	\includegraphics[width=0.7 \textwidth]{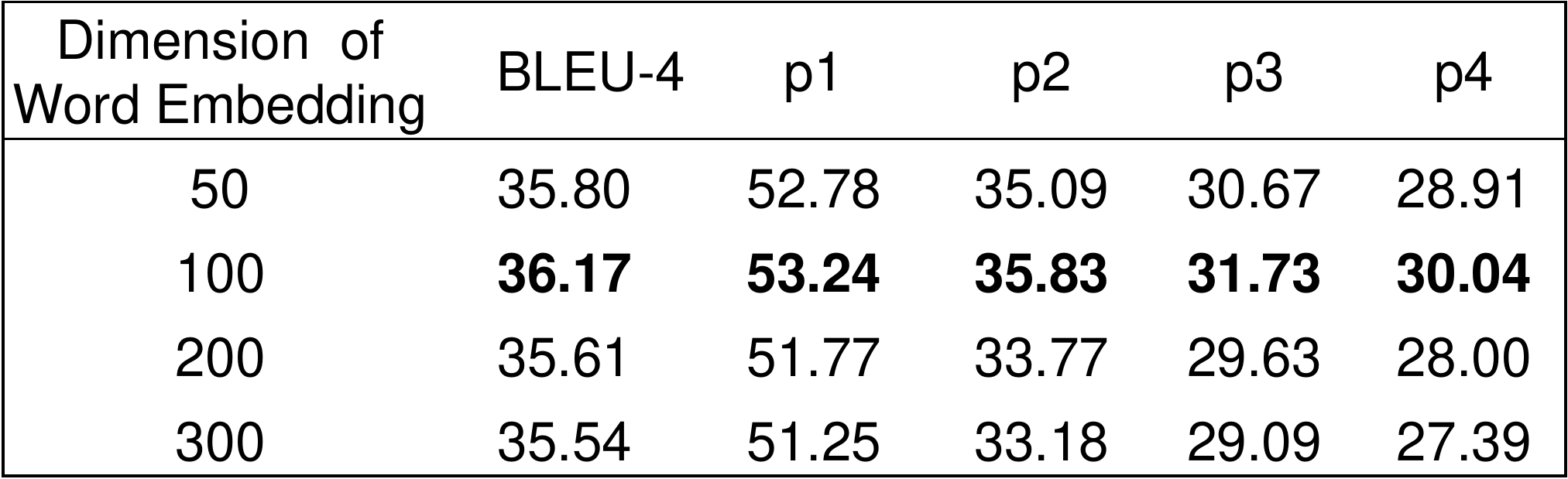}
	\vspace{-0.1cm}
	\caption{\small Different dimensions of word embedding.}
	\vspace{0.1cm}
    \end{subfigure}
    \begin{subfigure}[h]{0.24\textwidth}
    \centering
    \includegraphics[width=0.96 \textwidth]{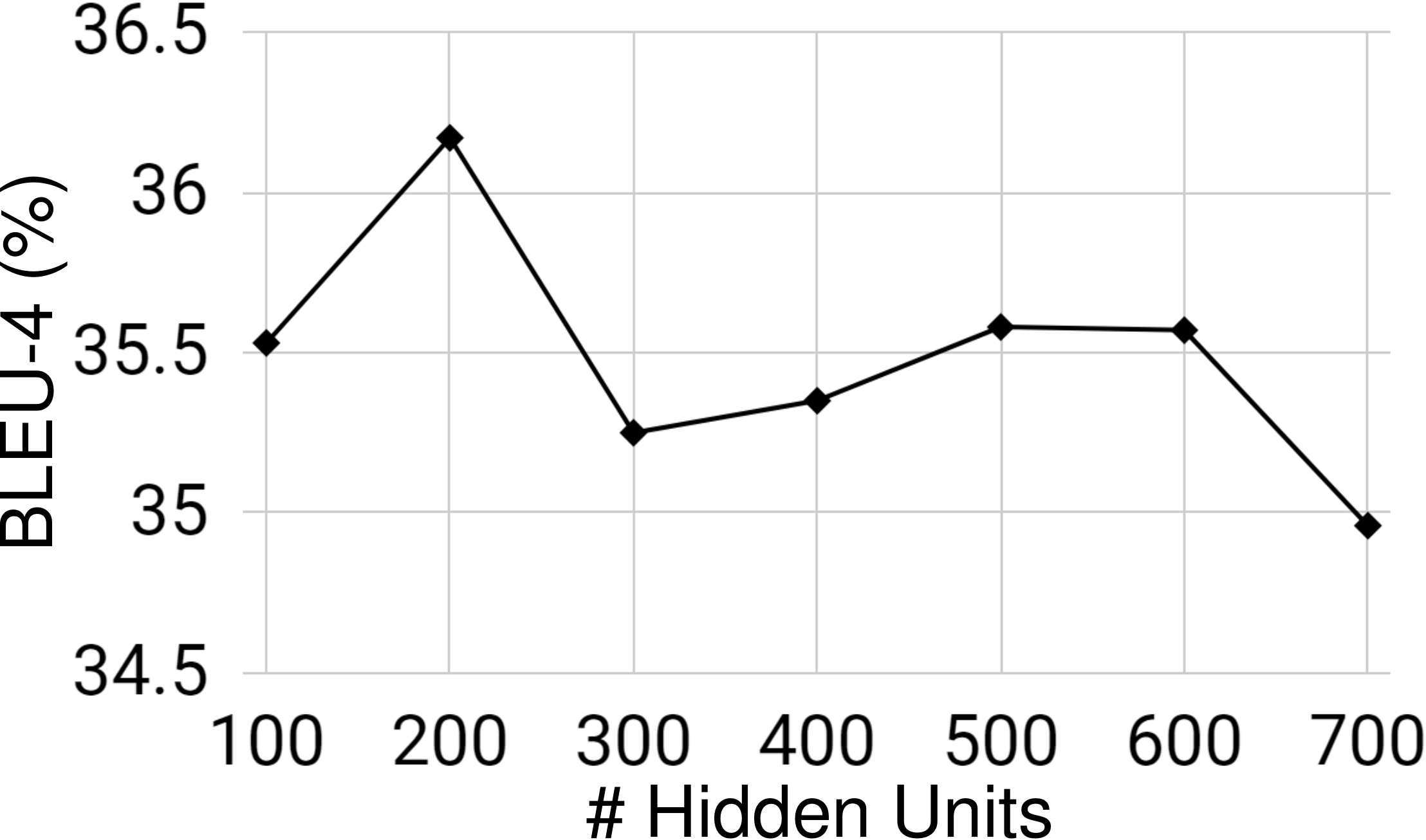}
	\caption{\small Different numbers of hidden units.}
    \end{subfigure}
    \begin{subfigure}[h]{0.24\textwidth}
    \centering
    \includegraphics[width=0.96 \textwidth]{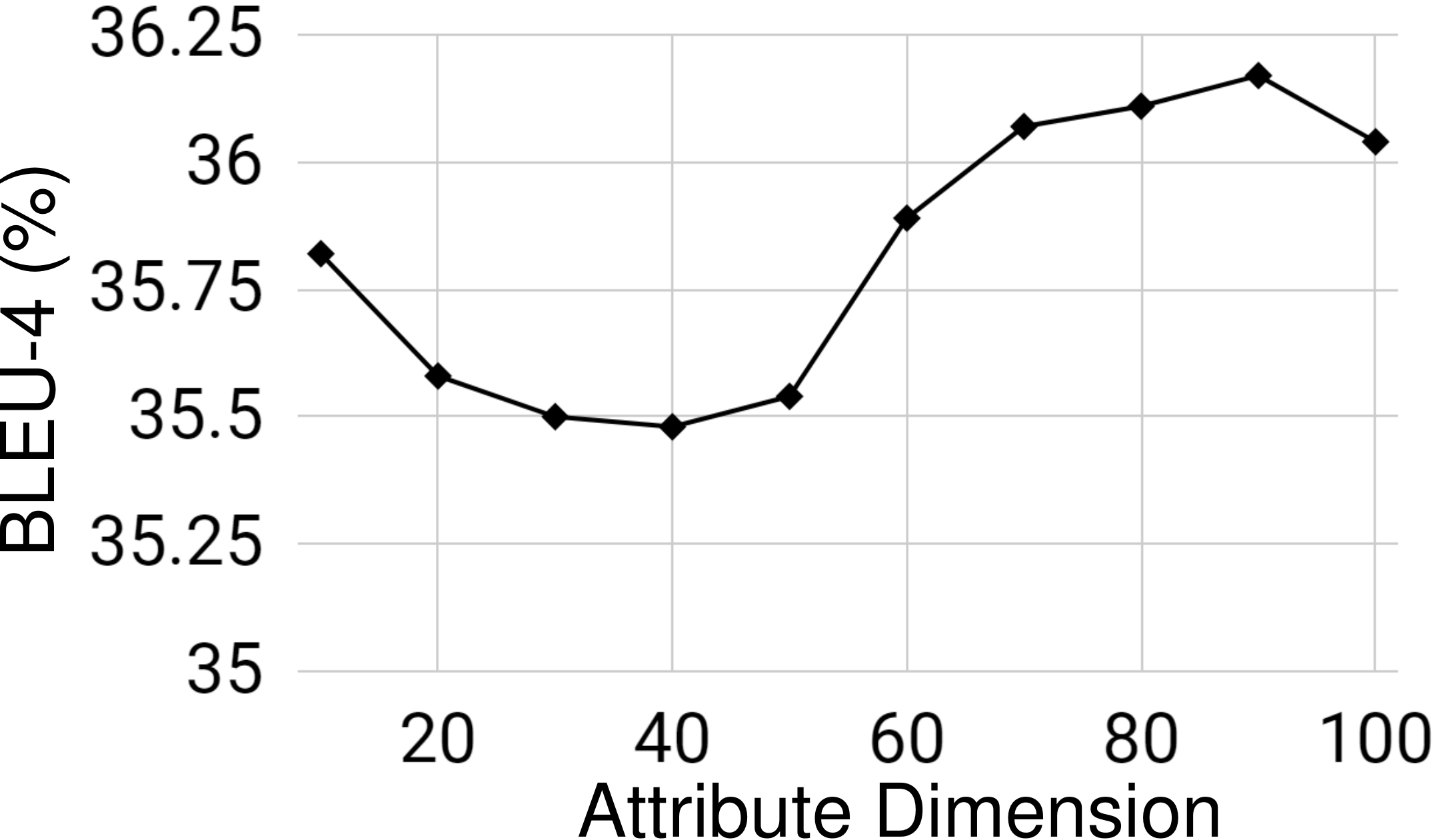}
	\caption{\small Different dimensions of component attribute embedding.}
    \end{subfigure}
    \caption{BLEU-4 scores of different parameter settings.}
	\label{fig:parameter}
\end{figure}

\subsection{RQ3: How accurate is RRGen under different parameter settings?}\label{subsec:param}

We also quantitatively compare the accuracy of RRGen in different parameter settings. We analyze three parameters, that is, the dimension of word embeddings, the number of hidden units, and also the dimension of component attribute embeddings. We vary the values of these three parameters and evaluate their impact on the BLEU-4 scores.

Figure~\ref{fig:parameter} shows the influence of different parameter settings on the test set. We choose the four different dimensions of word embeddings provided by GloVe~\cite{glove}, i.e., 50, 100, 200, and 300, and the result in Fig.~\ref{fig:parameter} (a) indicates that the RRGen model achieves the best BLEU-4 score when the word embedding size equals to 100. For the number of hidden units, we can see that more hidden units may not be helpful for improving accuracy, as shown in Fig.~\ref{fig:parameter} (b). RRGen generates the best result when we define the number of hidden units as 200. Fig.~\ref{fig:parameter} (c) shows that the accuracy of RRGen also changes along with the variations of attribute embedding dimension. The optimum dimension of attribute embedding is around 90.

\section{Human Evaluation}\label{sec:human}
In this section, we conduct a human evaluation to complement the evaluation in Section~\ref{sec:exper} that uses BLEU, since BLEU only measures the textual similarity between the generated responses and ground truth while the human study can evaluate users' general satisfaction on the responses. 

\subsection{Survey Procedure}
We conduct a human evaluation to evaluate the outputs of RRGen and compare RRGen with NMT and NNGen. We invite 20 participants, including 14 PhD students, two master students, one bachelor, and three senior researchers, all of whom are not co-authors and major in computer science. Among the participants, 15 of them have industrial experience in software development for at least a year, and eight of them have developed one or two mobile apps. Each participant is asked to read 25 user reviews, and assess the responses generated by NNGen, NMT, RRGen, and the app developers.

\subsection{Survey Design}
We randomly selected 100 review-response pairs in total, divide them evenly into four groups, and make a questionnaire for each group. We ensure that each review-response pair is evaluated by five different participants. In our questionnaire, each question presents the information of one review-response pair, i.e., its user review, the developer's response, its output from NNGen, and its responses generated by NMT and RRGen. The order of the responses from NNGen, NMT, RRGen, and official developers is randomly decided for each question.

Inspired by~\cite{DBLP:conf/acl/CardieD18,DBLP:conf/emnlp/LiS18}, all the response types are evaluated considering three aspects - ``\textit{grammatical fluency}'', ``\textit{relevance}'', and ``\textit{accuracy}''. We provided the following instructions at the beginning of each questionnaire to guide participants: The ``\textit{grammatical fluency}'' (or readability) measures the degree of whether a text is easy to understand; The metric ``\textit{relevance}'' relates to the extent of topical relevance between the user review and response; And the metric ``\textit{accuracy}'' estimates the degree of the response accurately answering a user review. 

\begin{figure}[h]
	\centering
	\includegraphics[width=0.48 \textwidth]{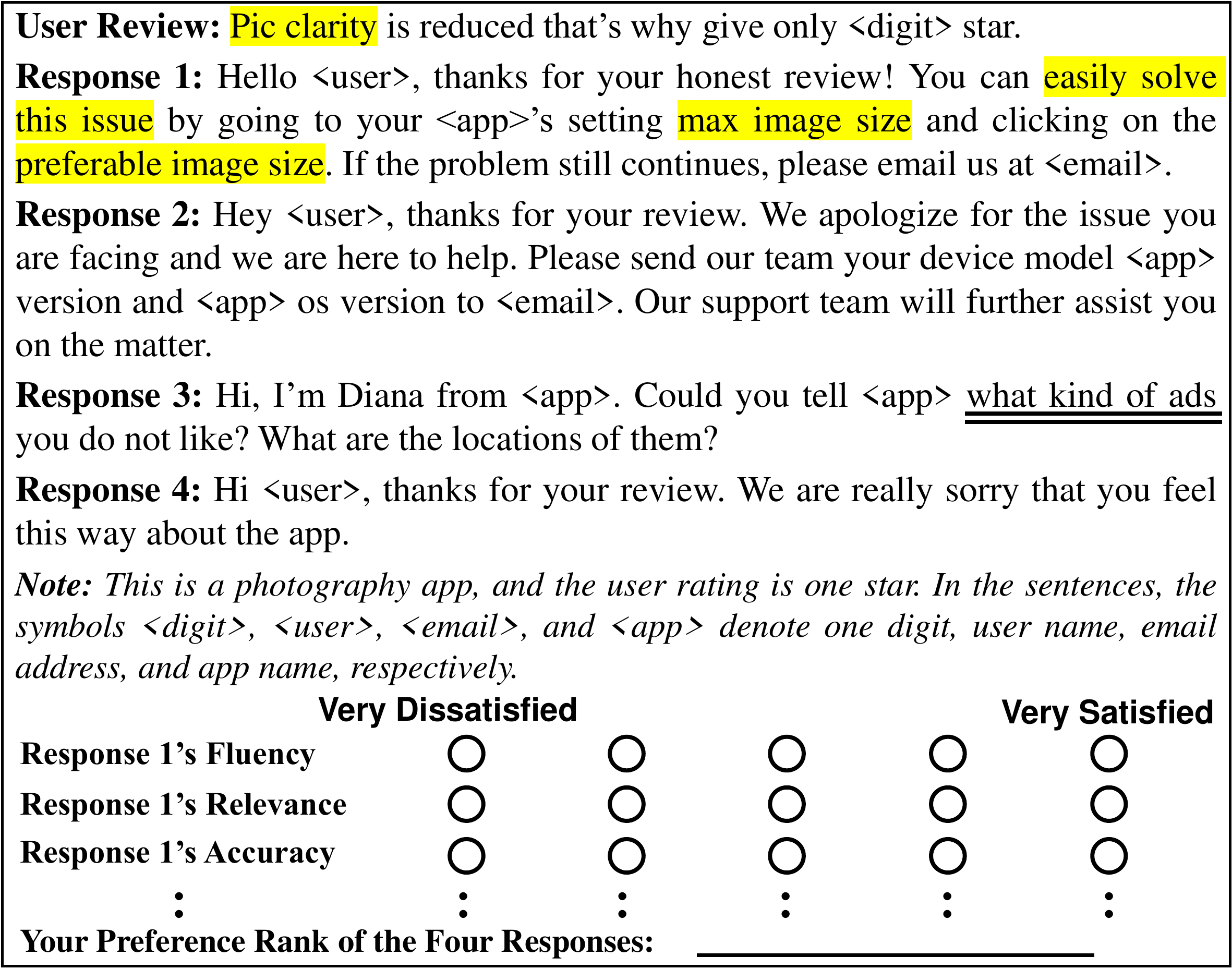}
	\vspace{0.1cm}
	\caption{A question in our survey. Response 1, 2, 3, and 4 correspond to the developer's response, the outputs of our RRGen model, and the responses produced by NNGen and NMT, respectively. Participants do not know the order of the four types of response during the survey, and are asked to score the three metrics for each response type. The two-dot symbols indicate the simplified grading schemes of Response 2, 3, and 4. The words highlighted in yellow are topical words in the descriptions, and the double-underlined words mean they are topically irrelevant to the user review.}
	\label{fig:question_example}
\end{figure}

All the three metrics are rated on a 1-5 scale (5 for fully satisfying the rating scheme, 1 for completely not satisfying the rating scheme, and 3 for the borderline cases), since a 5-point scale is widely used in prior software engineering studies~\cite{DBLP:conf/kbse/LiuXHLXW18,DBLP:conf/issta/KochharXLL16,di2016would}. Besides the three metrics, each participant is asked to rank responses generated by the three tools and those from developers based on their preference. The ``\textit{preference rank}'' score is rated on a 1-4 scale (1 for the most preferred). Fig.~\ref{fig:question_example} shows one question in our survey. Participants do not know which response is generated by which approach or whether it is written by developers, and they are asked to enter to score each response separately.

\subsection{Results}
We obtained 500 sets of scores from the human evaluation. Each set contains scores for the three metrics regarding the response of NNGen, NMT, RRGen, and official developers respectively, and also a ranking score of the four types of responses. The median time cost for one participant to complete his/her questionnaire is 0.94 hour, with an average value of 2.72 hours. We compute the agreement rate on the the preference ranks given by the participants, and find that 81\% of the total 100 review-response pairs received at least three identical preference ranks from the participants. Specifically, 31\%, 36\%, and 14\% were given the same preference ranks by three, four, and five participants respectively. This indicates that the participants achieved reasonable agreement on the performance of the generated responses.


Table~\ref{tab:human} shows the results of human evaluation. Bold indicates top scores. As expected, we can see that the response from official developers is preferred over the three approaches' outputs, which can be observed given the example in Fig.~\ref{fig:question_example}. Specifically, the developers' response (Response 1) is more relevant to the user review and provides more accurate solution to the app issue (e.g., reduced picture clarity) complained by the user. In terms of grammatical fluency, however, the RRGen model does quite well, achieving scores that are rather close to those of developers' responses, as shown in Table~\ref{tab:human}. In addition, we see that our RRGen model performs significantly better across all the metrics in comparison to the baseline approaches, which further indicates the effectiveness of RRGen in review response generation. 

\begin{table}[h]
	\center
	\caption{Human evaluation results for review response generation. Bold indicates top scores. Two-tailed t-test results are shown for our RRGen approach compared to NNGen and NMT (Statistical significance is indicated with \textsuperscript{*}($p-value<0.01$).).}
	\label{tab:human}
	\scalebox{0.9}{\begin{tabular}{l c c c c }
		\hline
		  & \begin{tabular}{@{}c@{}}Grammatical \\ Fluency\end{tabular} & Relevance & Accuracy & \begin{tabular}{@{}c@{}}Preference \\ Rank\end{tabular} \\
		 \hline
		 NNGen~\cite{DBLP:conf/kbse/LiuXHLXW18} & 4.520 & 3.160 & 3.104 & 3.339 \\
		 NMT~\cite{DBLP:journals/corr/BahdanauCB14} & 4.609 & 3.273 & 3.017 & 2.680 \\
		 RRGen & 4.626\textsuperscript{*} & 3.536\textsuperscript{*} & 3.458\textsuperscript{*} & 2.244\textsuperscript{*} \\
		 \hline
		 Developer & \textbf{4.644} & \textbf{3.804} & \textbf{3.712} & \textbf{1.736} \\
		\hline
	\end{tabular}
}
\end{table}
\vspace{-0.2cm}
\section{Discussion}\label{sec:dis}

\subsection{Why does Our Model Work?}
\begin{figure}[h]
    \begin{subfigure}[h]{0.5\textwidth}
    \centering
	\includegraphics[width=0.95 \textwidth]{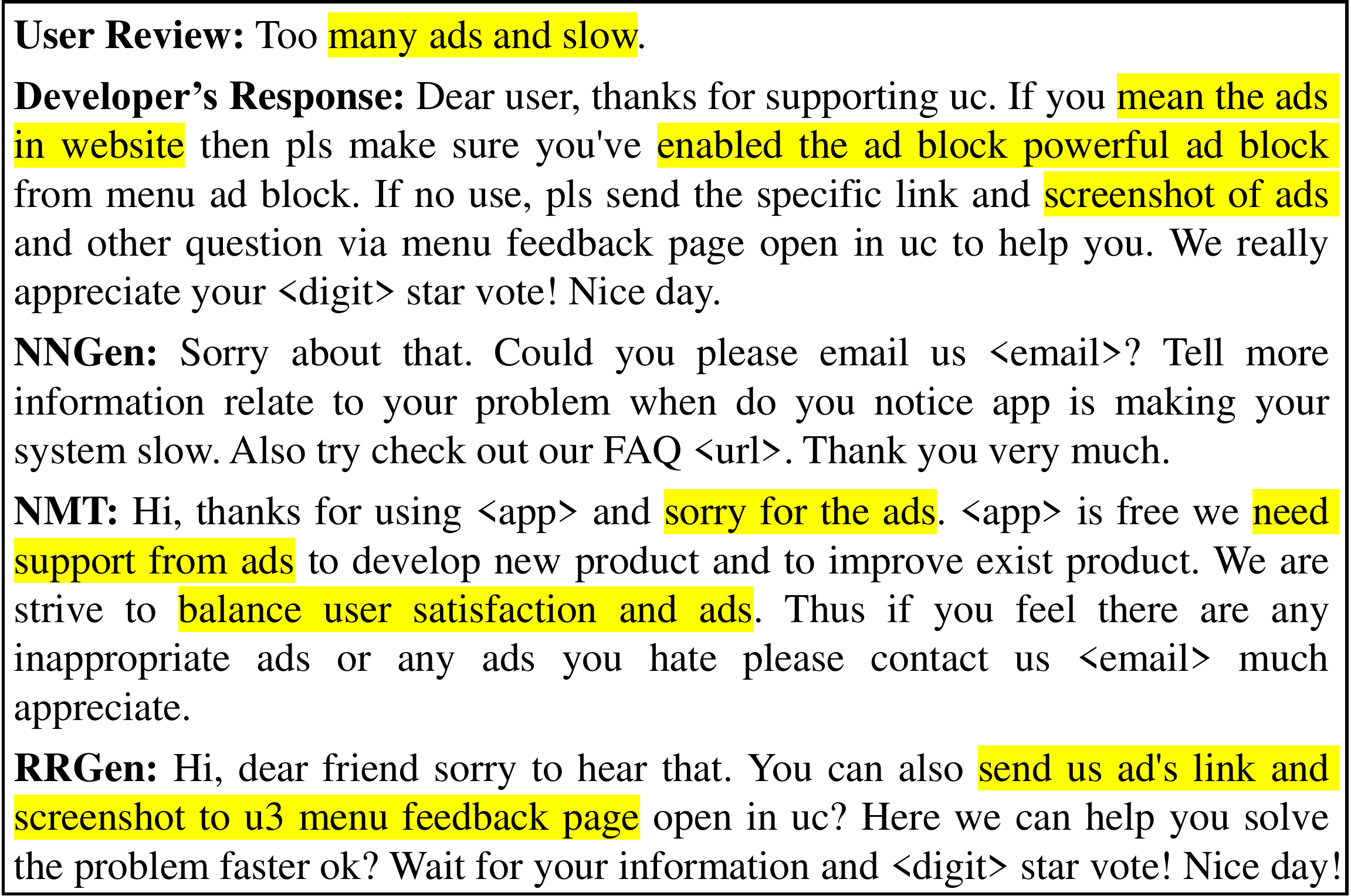}
	\captionsetup{aboveskip=1pt}
	\caption{\small First example.}
    \end{subfigure}
    
    \begin{subfigure}[h]{0.5\textwidth}
    \centering
    \includegraphics[width=0.95 \textwidth]{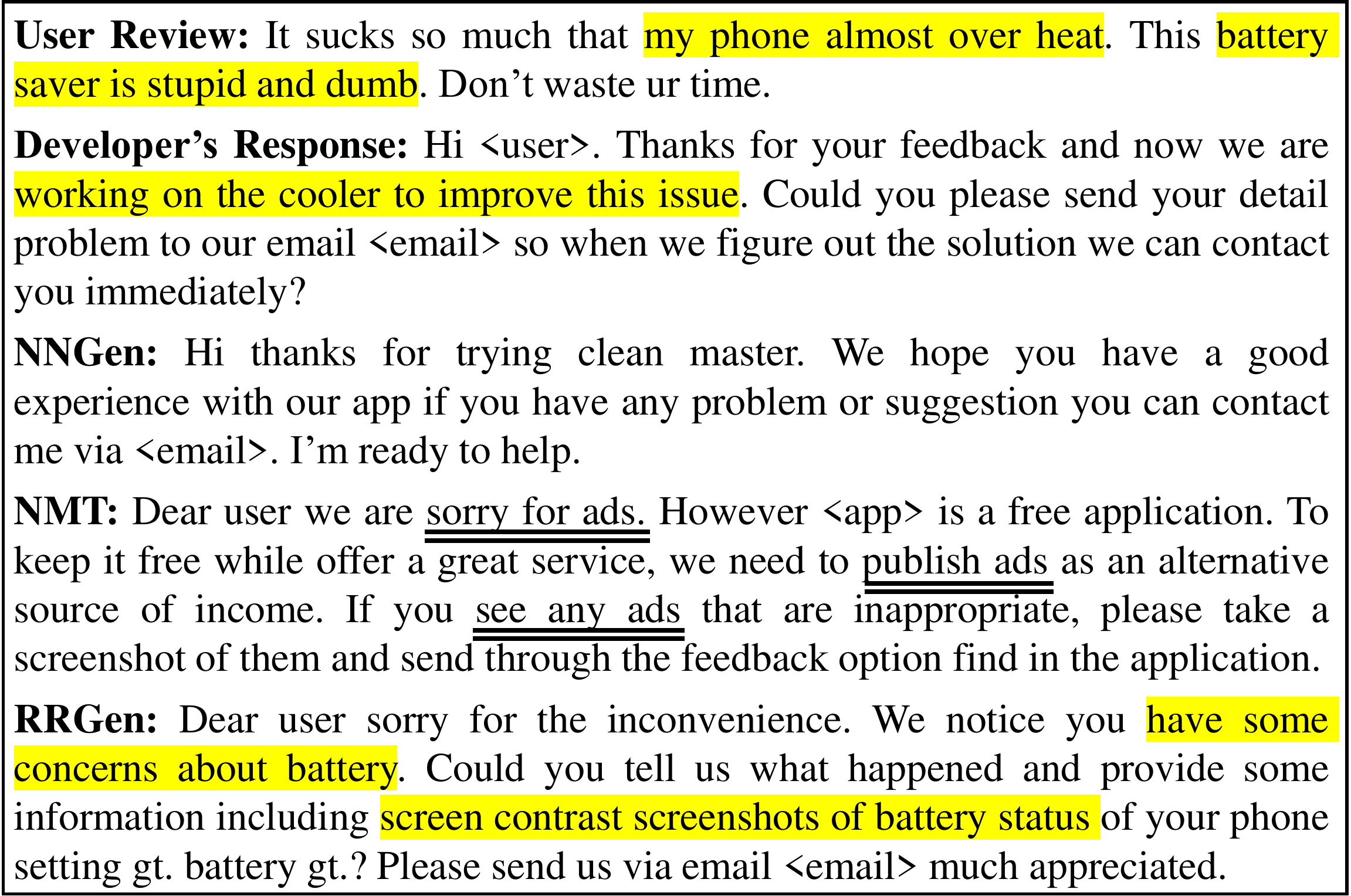}
    \captionsetup{aboveskip=1pt}
	\caption{\small Second example.}
    \end{subfigure}
    
    \caption{Two sample review-response pairs where RRGen can generate responses with more related topic. The meanings of the highlighted words and double-underlined words are the same as Fig.~\ref{fig:question_example}.}
	\label{fig:case}
\end{figure}
We have identified three advantages of RRGen that may explain its effectiveness in review response generation.

\textbf{Observation 1: RRGen can better capture salient topics of user reviews.} Unlike bag-of-words-based techniques, RRGen learns review and response representations with attentional deep learning. Characteristics of reviews, such as topical words and word orders, are naturally considered in these models~\cite{DBLP:conf/sigsoft/GuZZK16}. Moreover, keywords that are indicative of review topics are explicitly incorporated into the deep learning model, which would be helpful to better recognize the semantics of review and response. For example, it can learn that the review ``\textit{Too many ads and slow}'' is talking about the ad issue, and generate response related to the in-app ads, as shown in Fig.~\ref{fig:case} (a). In the example in Fig.~\ref{fig:case} (b), RRGen can well learn that the review is discussing about the battery issue, while NMT infers the topic wrongly. For the bag-of-words approach, NNGen, it may be easily confused by non-topical words. For the example in Fig.~\ref{fig:question_example}, NNGen (i.e., Response 3) focuses more on the words ``\textit{give}'', ``\textit{\textless digit\textgreater}'', and ``\textit{star}'', and selects the closest review ``\textit{If it has no advertising, I will give \textless digit\textgreater \,star}'' which has totally different topics comparing to the given review.

\begin{figure}[h]
    \begin{subfigure}[h]{0.5\textwidth}
    \centering
	\includegraphics[width=0.95 \textwidth]{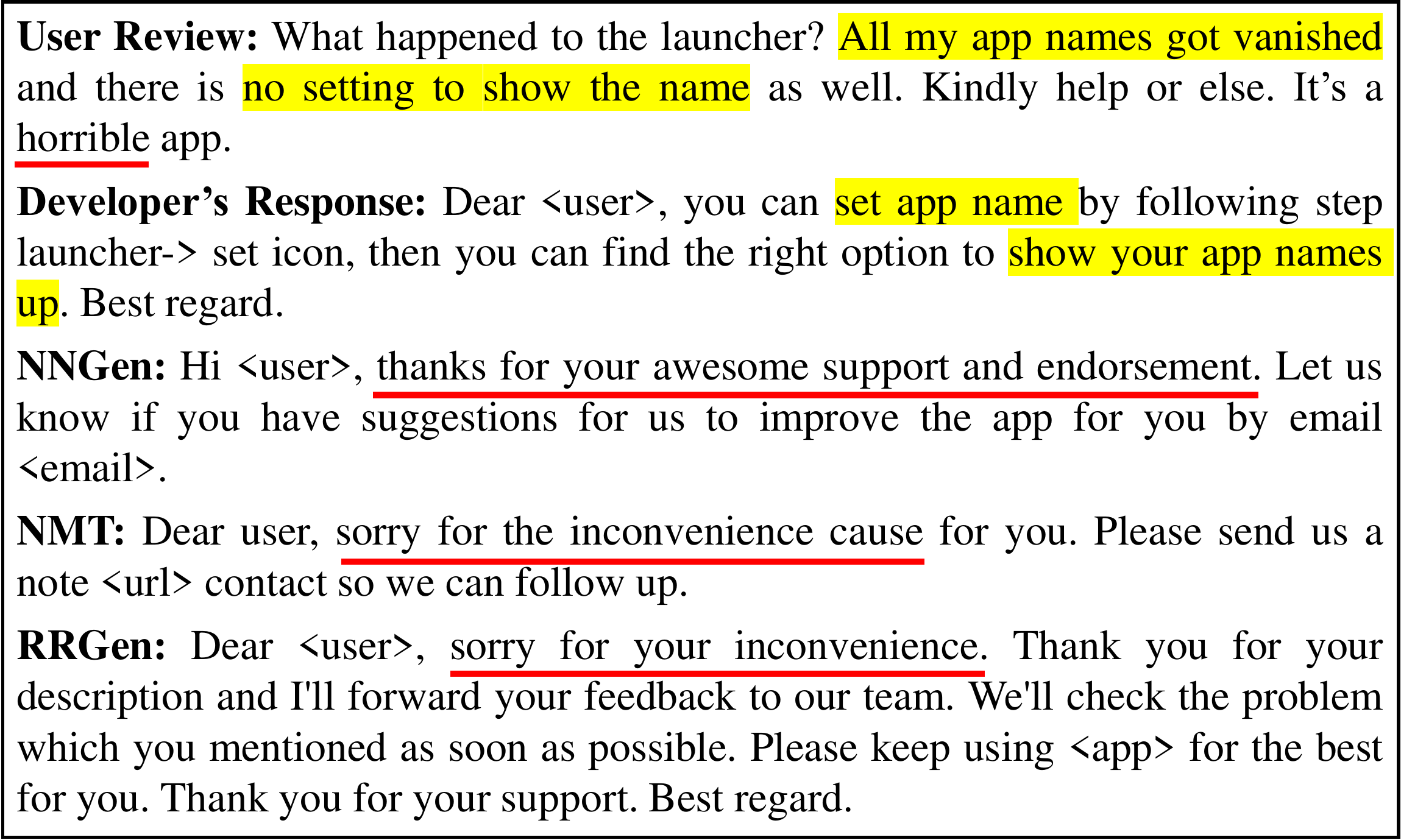}
	\captionsetup{aboveskip=1pt}
	\caption{\small First example.}
	\vspace{0.1cm}
    \end{subfigure}
    
    \begin{subfigure}[h]{0.5\textwidth}
    \centering
    \includegraphics[width=0.95 \textwidth]{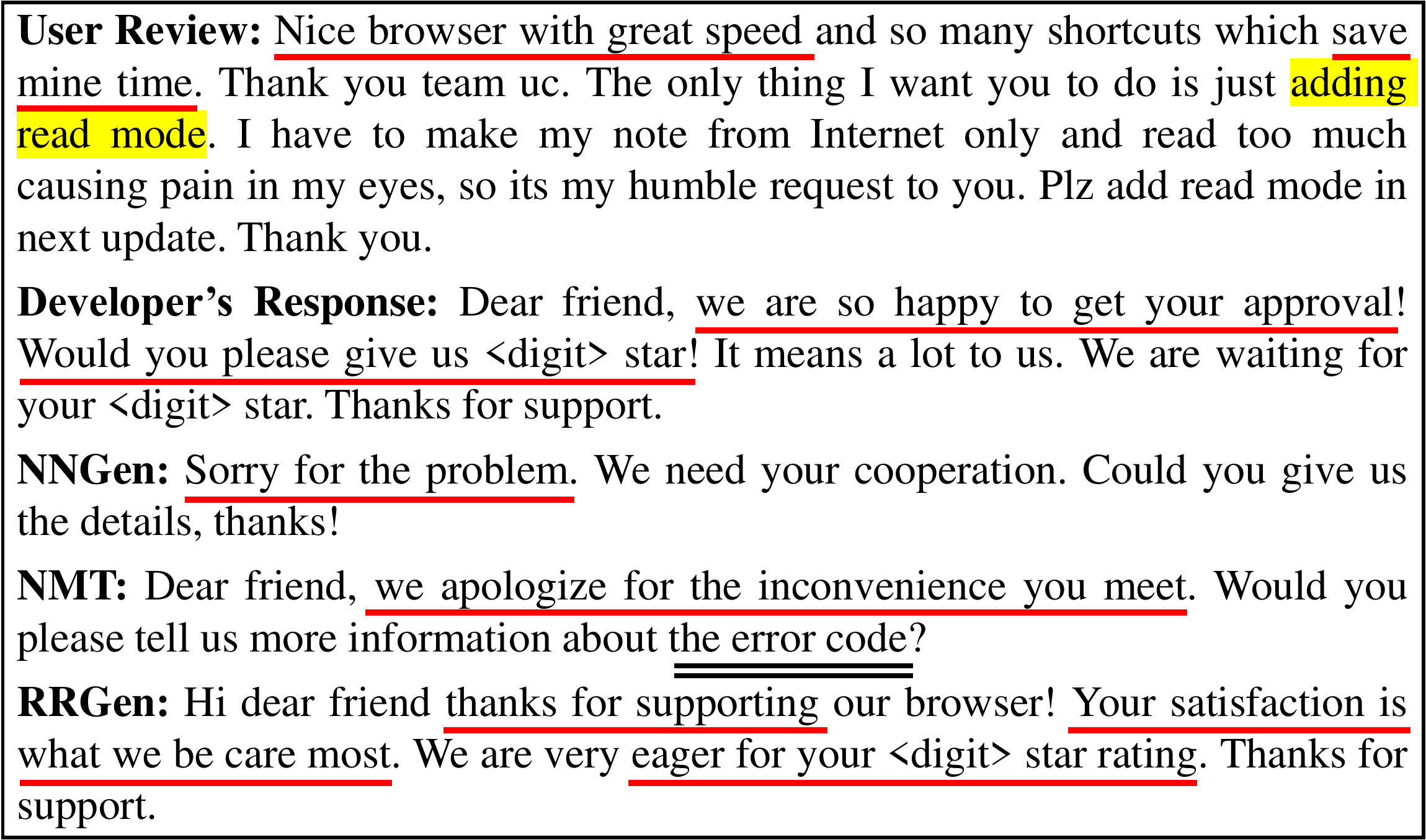}
    \captionsetup{aboveskip=1pt}
	\caption{\small Second example.}
    \end{subfigure}
    
    \caption{Two sample review-response pairs where RRGen can generate responses with more accurate sentiment. The words underlined in red are indicative of the description sentiments, and the meanings of the highlighted and double-underlined words are the same as Fig.~\ref{fig:question_example}.}
	\label{fig:case_sent}
\end{figure}

\textbf{Observation 2: RRGen can generate responses with more accurate sentiment.} User sentiment can be explicitly (e.g., the ``\textit{horrible}'' word in Fig.~\ref{fig:case_sent} (a)) or implicitly (e.g., the ``\textit{slow}'' word in Fig.~\ref{fig:case} (a)) reflected in user reviews. For the bag-of-words approach, the effect of sentiment words may be weakened by other words since their occurrence frequencies are similar. As can be seen in Fig.~\ref{fig:case_sent} (a), NNGen fails to infer the negative sentiment expressed by the review, and considers it as an endorsement message; while RRGen can accurately capture the negative information embedded in the review. Another example can be found in Fig.~\ref{fig:case_sent} (b), where both NNGen and NMT do not recognize that the positive sentiment of the given review. Without review attributes such as user ratings involved, NMT also fails to ask the user to increase his/her given rating.

\begin{figure*}[ht]
	\centering
	\includegraphics[width=0.94 \textwidth]{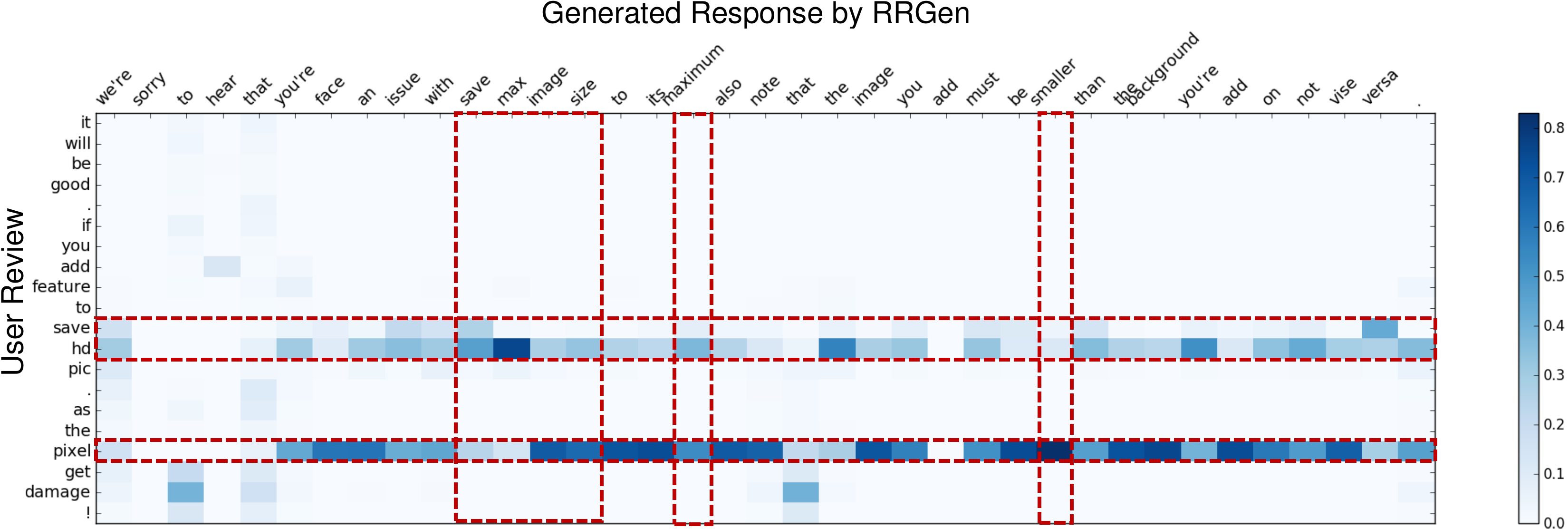}
	\caption{A heatmap representing the alignment between the user review (left) and generated response by RRGen (top). The columns represent the distribution over the user review after generating each word. Each pixel shows the weight $\alpha_{tj}$ of the annotation of the $j$-th source word for the $t$-th target word (see Equ.~(\ref{equ:attention})). A higher attention weight (indicated in darker color) manifests a stronger correlation between the target word and source word. The red dotted rectangles highlight partial topical words in corresponding descriptions.}
	\label{fig:attention_layer}
\end{figure*}

\textbf{Observation 3: RRGen can effectively capture knowledge relations between reviews and their corresponding responses.} RRGen learns the correspondence between reviews and response mainly through the high-dimensional hidden units and attention layer. The topical words in reviews tend to produce hidden states of semantically similar words in the RNN decoder. Fig.~\ref{fig:attention_layer} visualizes the latent alignment over the user review to help generate the response based on the attention weights $\alpha_{tj}$ from Equ.~(\ref{equ:attention}). Each column indicates the weight distribution over the user review for generating each word. From this we can see which words in the user review were considered more important when generating the target word in the response. We can observe the obvious correlations between the word ``\textit{save}'' (in the review) and ``\textit{save}'' (in the response), ``\textit{hd}'' (in the review) and ``\textit{max}'' (in the response), and ``\textit{pixel}'' (in the review) and ``\textit{image}'' (in the response), as shown in Fig.~\ref{fig:attention_layer}. This illustrates that RRGen is able to build implicit relations between the topical words in reviews and corresponding responses, which can help generate relevant and accurate response given a review.

\subsection{Post-Processing Steps}\label{subsec:post}

RRGen generates responses with placeholders, e.g., ``\textless email\textgreater'', ``\textless url\textgreater'', etc. Moreover, RRGen may not generate perfect responses and developers may want to verify RRGen responses for some more ``sensitive'' cases. To partially address the above-mentioned limitations, we propose several post-processing steps. First, we build a placeholder-value dictionary for automatically replacing placeholders (e.g., ``\textless url\textgreater'') with corresponding values (e.g., ``https://www.facebook.com/groups/vivavideoapp/'') for each app. Second, we design a quality assurance filter to automatically detect the generated responses that \textit{require further check}.

The placeholder-value dictionary for each app is saved during preprocessing, and for simplicity, only the most common value for each placeholder is saved. We define a generated response requiring further check based on its token length $l$, the overlapped keyword ratio $\omega$ with the corresponding review, and also the review rating $r$. Specifically, we define responses that satisfy the following constraint, i.e., $\omega<0.05$ or $(l<38 \: and \: r\leq 2)$ to require further check. The thresholds are determined as follows: 0.05 is determined by following the keyword overlapping threshold in~\cite{di2016would}, 38 is the first quartile of response token lengths in the whole dataset, and the constraint for review rating is set as such as reviews with lower ratings (e.g., 1, 2) tend to express users' strong dissatisfaction with certain aspects of apps~\cite{chen2014ar,DBLP:conf/sose/GaoXHZ15}. 

We evaluate our solution after the above mentioned post-processing strategy using a similar experiment setting used to produce results presented in Section~\ref{subsec:accuracy}. We find that the BLEU-4 score is 34.63. It is only slightly lower than the BLEU-4 score (36.17) reported in our earlier experiment using ground truths with placeholders rather than actual values.

\subsection{Limitations}
Although our proposed RRGen model aims at producing accurate responses to user reviews, not all the reviews require responses, some reviews require carefully crafted replies, and some other reviews can be delegated to an automated bot. We have tried to address this issue partially by adding some preliminary post-processing steps (see Section~\ref{subsec:post}).


Admittedly, our post-processing steps are not perfect. First, our preliminary post-processing steps may generate responses with inappropriate values due to the coarsely-defined placeholder-value dictionary. This issue can be improved by creating a context-sensitive dictionary for each app. Also, our simple rule-based detection of responses that require further check can be improved further. For this, we can learn the thresholds of the rule conditions or design new detection criteria. We leave the design, implementation, and evaluation of a full-fledged system that can route reviews to {\em do not respond}, {\em require human careful response}, and {\em can be responded by an automated bot} queues for future work. As our work is the first to automate app review generation, although it is not perfect, it opens up way for future research to continue our study and improve it further.

\subsection{Threats to Validity}
One of the threats to validity is about the limited number of studied apps. We studied developer responses for reviews of free apps only. One of the main reasons for removing non-free apps is that the pricing of an app is likely to impact developers' response behavior~\cite{DBLP:journals/ese/HassanTBH18}. Also, we only consider Google Play apps in this work, because Apple's App Store started to support review response from 2017 while the feature has been standard in Google Play since 2013~\cite{iosvsgoogleresponse}. Although our study is based on apps from various categories and large numbers of review-response pairs, future work can be extended to multiple app stores and paid apps.

The second threat to validity is about the component attributes incorporated into our proposed model. Although we involve both high-level attributes and keywords, some other characteristics such as review title length and post date, which would be helpful for response generation, are not considered. Besides, the review sentiment predicted by SentiStrength~\cite{DBLP:journals/jasis/ThelwallBPCK10} might not be reliable~\cite{DBLP:conf/msr/NovielliGL08}, and could influence the generated response. However, accurate sentiment prediction based on reviews is out of the scope of this paper, and the effectiveness of StentiStrength in detecting user sentiment about app features has been demonstrated in~\cite{guzman2014users}. In the future, we will explore the impact of more review characteristics on automatic review response generation.

Another threat to validity is about manual inspection in Section~\ref{sec:human}. The results of the human evaluation are impacted by the experience of the participants and their intuition of the evaluation metrics. To reduce the errors in the manual analysis, we ensure that each review-response pair was evaluated by five different participants. As our participants are mainly students, they may not be representative of (CRM) professionals who are likely to benefit from our tools in practice~\cite{DBLP:conf/icse/SalmanMJ15,DBLP:journals/ese/FeldtZBFJJMORSS18}. We try to mitigate this threat by inviting the students with at least one year of software development experience. In addition, we randomly disrupt the order of the three types of response for each question, so that the results are not influenced by participants' prior knowledge about the response orders.

\section{Related Work}\label{sec:related}

\subsection{User Review Mining}

Identifying the complaint topics expressed by user reviews is the basis for user review mining~\cite{Palomba2017ICSE,Grano2018,DBLP:conf/icse/GaoZD0ZLK19}. Iacob et al.~\cite{DBLP:conf/bcshci/IacobVH13} manually label 3,278 reviews, and discover the most recurring issues users report through reviews. To alleviate the labor in manual labeling, many studies focus on automating the process. For example, Iacob and Harrison \cite{DBLP:conf/msr/IacobH13} design MARA for retrieving app feature requests based on linguistic rules. Maalej and Nabil~\cite{DBLP:conf/re/MaalejN15} adopt probabilistic techniques to classify reviews. Di Sorbo et al.~\cite{di2016would} separately categorize user intentions and topics delivered by app reviews. Understanding user sentiment about specific app aspects is another typical direction of review mining. Guzman and Maalej~\cite{guzman2014users} use topic modeling approach and StentiStrength~\cite{DBLP:journals/jasis/ThelwallBPCK10} (a lexical sentiment extraction tool) to predict sentiment of app features. Gu and Kim~\cite{DBLP:conf/kbse/GuK15} propose SUR-Miner to exploit grammatical structures for aspect-opinion identification. More research of mobile review analysis can be found in~\cite{DBLP:journals/tse/MartinSJZH17}. Different from these existing review analysis research, we contribute to facilitating the bidirectional dialogue between users and developers instead of analyzing only the feedback from user side.



\subsection{Analysis of User-Developer Dialogues in App Stores}
Oh et al.~\cite{DBLP:conf/chi/OhKLLS13} conduct a survey on 100 smartphone users to understand how developers and users interact. They find that most users (69\%) tend to take a passive action such as uninstalling apps, and the main reason for such behavior is that these users think that their inquiries (e.g., user reviews) would take long time to be responded or receive no response. McIlroy et al.~\cite{DBLP:journals/software/McIlroySAH17} analyze reviews of 10,000+ free Google Play apps and find that 13.8\% of the apps respond to at least one review. They also observe that users would change their ratings 38.7\% of the time following a response. Such positive impact of developers' response is also confirmed by Hassan et al.~\cite{DBLP:journals/ese/HassanTBH18}. Although these studies do highlight the importance of responding to user reviews, they do not provide an explicit method to alleviate the burden in the responding process, which is the focus of this work.

\subsection{Short Text Dialogue Analysis}
Short text dialogue analysis is one popular topic in the field of natural language processing, in which given a message from human, the computer returns a reasonable response to the message~\cite{DBLP:journals/corr/JiLL14,DBLP:journals/tacl/ZengLHGLK19}. Short text dialogue can be formalized as a search or a generation problem. The former formalization is based on a knowledge base consisting of a large number of message-response pairs. Information retrieval techniques~\cite{DBLP:books/daglib/0021593} are generally utilized to select the most suitable response to the current message from the knowledge base. The major bottleneck for search-based approaches is the creation of the knowledge base~\cite{DBLP:conf/flairs/ChenTALT11}. Ritter et al.~\cite{DBLP:conf/emnlp/RitterCD11} and Vinyals and Le~\cite{DBLP:journals/corr/VinyalsL15} are the first to treat generation of conversational dialog as a data-driven statistical machine translation (SMT)~\cite{DBLP:conf/acl/KoehnHBCFBCSMZDBCH07} problem. Their results show that the machine translation-based approach works better than one IR approach, vector space model (VSM)~\cite{DBLP:journals/cacm/SaltonWY75}, in terms of BLEU score~\cite{DBLP:conf/acl/PapineniRWZ02}. However, generation-based approaches cannot guarantee that the response is a legitimate natural language text. In this work, we propose to integrate app reviews' unique characteristics for accurate response generation.


\section{Conclusion and Future Work}\label{sec:con}
Replying to user reviews can help app developers create a better user experience and improve apps' ratings. Due to the large numbers of reviews received for popular apps each day, automating the review response process is useful for app developers. In this work, we propose a novel approach named RRGen by explicitly incorporating review attributes and occurrences of specific keywords into the basic NMT model. Analysis using automated metric and human evaluation shows that our proposed model outperforms baseline approaches. In future, we will conduct evaluation using a larger dataset and deploy the model with our industry partners.


\section*{Acknowledgement}
The work described in this paper was supported by the Research Grants Council of the Hong Kong Special Administrative Region, China (No. CUHK 14210717 and No. CUHK 14208815 of the General Research Fund), and Microsoft Research Asia (2018 Microsoft Research Asia Collaborative Research Award).

\bibliographystyle{IEEEtran}
\bibliography{main}

\end{document}